\documentclass[]{aastex63}

\let\tablenum\relax
\usepackage{siunitx}
\usepackage[caption=false]{subfig}

\usepackage{booktabs}
\usepackage{makecell}
\usepackage[version=4]{mhchem}

\usepackage{epigraph}
\usepackage[T1]{fontenc}

\received{}
\revised{}
\accepted{}
\submitjournal{PSJ}

\shorttitle{THAI Part 1}
\shortauthors{Turbet et al.}

\graphicspath{{./}{figures_part1/}}

\begin{document}

\title{The TRAPPIST-1 Habitable Atmosphere Intercomparison (THAI). \\
 Part I: Dry Cases -- The fellowship of the GCMs}

\correspondingauthor{Martin Turbet}
\email{martin.turbet@lmd.ipsl.fr}
\author[0000-0003-2260-9856]{Martin Turbet}
\affiliation{Laboratoire de M\'et\'eorologie Dynamique/IPSL, CNRS, Sorbonne Universit\'e, \'Ecole Normale Sup\'erieure, PSL Research University, \'Ecole Polytechnique, 75005 Paris, France}
\affiliation{Observatoire  Astronomique  de  l’Universit\'e  de  Gen\`eve,  Universit\'e  de  Gen\`eve,  Chemin  des Maillettes 51, 1290 Versoix, Switzerland.}

\author[0000-0002-5967-9631]{Thomas J. Fauchez}
\affiliation{NASA Goddard Space Flight Center,
8800 Greenbelt Road,
Greenbelt, MD 20771, USA}
\affiliation{Goddard Earth Sciences Technology and Research (GESTAR), Universities Space Research Association (USRA), Columbia, MD 7178, USA}
\affiliation{NASA GSFC Sellers Exoplanet Environments Collaboration}

\author[0000-0001-8832-5288]{Denis E. Sergeev}
\affiliation{Department of Mathematics,
College of Engineering, Mathematics, and Physical Sciences, University of Exeter,
Exeter, EX4 4QF, UK}

\author[0000-0002-1485-4475]{Ian A. Boutle}
\affiliation{Met Office, FitzRoy Road, Exeter, EX1 3PB, UK}
\affiliation{Department of Astrophysics, College of Engineering, Mathematics, and Physical Sciences, University of Exeter, Exeter, EX4 4QL, UK}

\author[0000-0001-5328-819X]{Kostas Tsigaridis}
\affiliation{Center for Climate Systems Research, Columbia University, New York, NY, USA}
\affiliation{NASA Goddard Institute for Space Studies, 2880 Broadway, New York, NY 10025, USA}

\author[0000-0003-3728-0475]{Michael J. Way}
\affiliation{NASA Goddard Institute for Space Studies, 2880 Broadway, New York, NY 10025, USA}
\affiliation{NASA GSFC Sellers Exoplanet Environments Collaboration}
\affiliation{Theoretical Astrophysics, Department of Physics and Astronomy, Uppsala University, Uppsala, Sweden}

\author[0000-0002-7188-1648]{Eric T. Wolf}
\affiliation{Laboratory for Atmospheric and Space Physics, University of Colorado Boulder, Boulder, CO, USA}
\affiliation{NASA NExSS Virtual Planetary Laboratory, Seattle, WA, 98195, USA}
\affiliation{NASA GSFC Sellers Exoplanet Environments Collaboration}

\author[0000-0003-0354-9325]{Shawn D. Domagal-Goldman}
\affiliation{NASA Goddard Space Flight Center, 8800 Greenbelt Road, Greenbelt, MD 20771, USA}
\affiliation{NASA GSFC Sellers Exoplanet Environments Collaboration}
\affiliation{NASA NExSS Virtual Planetary Laboratory, Seattle, WA, 98195, USA}

\author[0000-0002-3262-4366]{Fran\c cois Forget}
\affiliation{Laboratoire de M\'et\'eorologie Dynamique/IPSL, CNRS, Sorbonne Universit\'e, \'Ecole Normale Sup\'erieure, PSL Research University, \'Ecole Polytechnique, 75005 Paris, France}

\author[0000-0003-4346-2611]{Jacob Haqq-Misra}
\affiliation{NASA NExSS Virtual Planetary Laboratory, Seattle, WA, 98195, USA}
\affiliation{Blue Marble Space Institute of Science, Seattle, WA, USA}

\author[0000-0002-5893-2471]{Ravi K. Kopparapu}
\affiliation{NASA Goddard Space Flight Center, 8800 Greenbelt Road, Greenbelt, MD 20771, USA}
\affiliation{NASA GSFC Sellers Exoplanet Environments Collaboration}
\affiliation{NASA NExSS Virtual Planetary Laboratory, Seattle, WA, 98195, USA}

\author[0000-0002-4664-1327]{F. Hugo Lambert}
\affiliation{Department of Mathematics, College of Engineering, Mathematics, and Physical Sciences, University of Exeter, Exeter, EX4 4QF, UK}

\author[0000-0003-4402-6811]{James Manners}
\affiliation{Met Office, FitzRoy Road, Exeter, EX1 3PB, UK}

\author[0000-0001-6707-4563]{Nathan J. Mayne}
\affiliation{Department of Astrophysics, College of Engineering, Mathematics, and Physical Sciences, University of Exeter, Exeter, EX4 4QL, UK}

\author[0000-0002-6673-2007]{Linda Sohl}
\affiliation{NASA Goddard Institute for Space Studies, 2880 Broadway, New York, NY 10025, USA}
\affiliation{Center for Climate Systems Research, Columbia University, New York, NY, USA}

\begin{abstract}

With the commissioning of powerful, new-generation telescopes such as the JWST and the ground-based ELTs, the first characterization of a high-molecular-weight atmosphere around a temperate rocky exoplanet is imminent. Atmospheric simulations and synthetic observables of target exoplanets are essential to prepare and interpret these observations. Here we report the results of the first part of the THAI (TRAPPIST-1 Habitable Atmosphere Intercomparison) project, which compares 3D numerical simulations performed with four state-of-the-art Global Climate Models (ExoCAM, LMD-Generic, ROCKE-3D, Unified Model) for the potentially habitable target TRAPPIST-1e. In this first part, we present the results of dry atmospheric simulations. These simulations serve as a benchmark to test how radiative transfer, subgrid-scale mixing (dry turbulence and convection) and large-scale dynamics impact the climate of TRAPPIST-1e and consequently the transit spectroscopy signature as seen by JWST. To first order, the four models give results in good agreement. The inter-model spread in the global mean surface temperature amounts to 7K (6K) for the N$_2$-dominated (CO$_2$-dominated, respectively) atmosphere. The radiative fluxes are also remarkably similar (inter-model variations less than 5$\%$), from the surface (1bar) up to atmospheric pressures $\sim$5millibar. Moderate differences between the models appear in the atmospheric circulation pattern (winds) and the (stratospheric) thermal structure. These differences arise between the models from (1) large scale dynamics because TRAPPIST-1e lies at the tipping point between two different circulation regimes (fast and Rhines rotators) in which the models can be alternatively trapped; and (2) parameterizations used in the upper atmosphere such as numerical damping.

\end{abstract}



\section{Introduction} \label{sec:intro}
\epigraph{Don't adventures ever have an end? I suppose not. Someone else always has to carry on the story.}{---J.R.R. Tolkien, \textit{The Fellowship of the Ring}}


With the commissioning of next-generation telescopes and instruments, the astronomical community is for the first time in the process of detecting and characterizing the chemical composition of the atmospheres of rocky exoplanets receiving a moderate instellation \citep{Greene:2016,Morley2017,Wunderlich2019,gillon2020trappist1,Turbet:2020trappist1review}. Specifically, observations of the TRAPPIST-1 planets \citep{Gillon2017} with the James Webb Space Telescope (JWST) are likely to be our first real opportunity to characterize the surface and atmosphere of planets with a similar mass, radius and bolometric irradiation to Earth but orbiting other stars than the Sun \citep{Morley2017,Lustig_Yaeger2019,Fauchez:2019,Wunderlich2019,Wunderlich2020,gillon2020trappist1,Turbet:2020trappist1review}.

Preparing for these observations with sophisticated numerical climate models has recently become an active topic of research aimed at optimising the scientific return from observational campaigns (see \citealt{Fauchez2021_THAI_workshop} and references therein). Firstly, model predictions guide observation proposals, and secondly they are used to help in the analysis and interpretation of the data. However, little work has been done so far to identify sources of uncertainty between atmospheric models employed in the exoplanet community, and how this may affect the feasibility as well as the interpretation of observations of exoplanet atmospheres. To this end, global climate model intercomparisons projects, which have been pioneered and successfully used to study the past, present and future evolution of the Earth's climate \citep[e.g.][]{Eyring16}, have also been used more recently in the context of exoplanets \citep[e.g.][]{Yang2019}.

The TRAPPIST-1 Habitable Atmosphere Intercomparison (THAI) project \citep{Fauchez2020THAI} brings together four state-of-the-art 3D Global Climate Models to simulate the climate and observability with JWST of TRAPPIST-1e under various atmospheric composition scenarios. TRAPPIST-1e is to date the best potentially habitable exoplanet target we have \citep{Wolf2017,Turbet:2018aa,Fauchez:2019,Sergeev2020}. As of September 2021, 4 transit observations of TRAPPIST-1e are already planned in the first observation cycle of JWST using the NIRSpec instrument (GTO Program 1331, PI: Nikole Lewis).

The goal of the THAI intercomparison project is twofold. The first aim is to identify the extent to which, for given atmospheric composition scenarios, the different models agree or disagree on the representation of the climate of TRAPPIST-1e. This first investigation is key to identifying the processes responsible for most of the disparities between models, and therefore where the community must deploy significant effort.
The second aim is to assess how uncertainties in climate modelling affect observables, which provides a degree of confidence in our interpretation of observations of the atmospheric composition and climate of TRAPPIST-1e.

The results of the THAI intercomparison project are presented in three parts detailed in three distinct manuscripts. In the first part (this manuscript), we present and compare the results of dry (i.e. without water cycle and without clouds) 3D numerical climate simulations of TRAPPIST-1e performed with four GCMs (ExoCAM, LMD-Generic,  ROCKE-3D, UM). These simulations are used as benchmark cases to better understand the differences in the water cycle (moist convection, clouds) described in \citet[][referred to as Part 2 of THAI]{Sergeev21_THAI} and their effects on the synthetic observations described in \citet[][referred to as Part 3 of THAI]{Fauchez21_THAI}. The benchmark simulations presented here were designed to test the impact of the radiative transfer (RT), the dynamical core, the boundary layer parameterization and the sub-grid scale dry dynamics on the mean climate state and ultimately on the synthetic observables of TRAPPIST-1e \citep{Fauchez21_THAI}.

The manuscript is structured as follows: In Section~\ref{sec:method} the methods and tools used in this study are presented, with particular emphasis on the description of the parameterizations (RT, dynamical core, sub-grid dynamics) used. In Section~\ref{sec:results}, we present the results of this first intercomparison phase on the climate states of TRAPPIST-1e and discuss the possible source of differences between the models. Finally, conclusions are given in section~\ref{sec:conclusions}.

\section{Methods} \label{sec:method}

\subsection{Planetary configuration}
\begin{deluxetable*}{lll}
\tablecaption{TRAPPIST-1 stellar spectrum and planetary parameters of TRAPPIST-1e \citep{Grimm18} used for the THAI simulations. \label{tab:planet}}
\tablehead{
\colhead{Parameter} & \colhead{Units} & \colhead{Value}
}
\startdata
Star and spectrum &                                & \SI{2600}{\K} BT-Settl with Fe/H=0 \\
Semi-major axis   & AU                             & 0.02928 \\
Orbital period    & Earth day                      & 6.1 \\
Rotation period   & Earth day                      & 6.1 \\
Obliquity         &                                & 0 \\
Eccentricity      &                                & 0 \\
Instellation      & \si{\watt\per\square\meter}    & 900.0 \\
Planet radius     & \si{\km}                       & 5798 \\
Gravity           & \si{\meter\per\second\squared} & 9.12 \\
\enddata
\end{deluxetable*}

In this Part 1 of the THAI trilogy, we present results of the dry atmosphere simulations, named Ben~1 \& Ben~2 in the THAI protocol \citep{Fauchez2020THAI}.
The Ben~1 \& Ben~2 cases were designed to act as benchmarks for the two habitable states Hab~1 \& Hab~2, respectively, of TRAPPIST-1e, described in Part 2 \citep{Sergeev21_THAI}. Ben~1 is the colder state with a 1~bar \ce{N2}-dominated atmosphere plus 400 ppm of \ce{CO2}, while Ben~2 is a warmer state with a 1~bar \ce{CO2}-dominated atmosphere.

In these simulations, the planet is assumed to be tidally locked to its host star TRAPPIST-1. Both simulations are started from an isothermal (\SI{300}{\K}) dry atmosphere at rest. The models are integrated until top-of-atmosphere (TOA) radiative balance is reached. Model outputs are then provided for 10 orbits (i.e., 61 Earth days) and every \SI{6}{\hour}. Readers can refer to \citet{Fauchez2020THAI} for more details on the experimental setup of the Ben~1 \& Ben~2 simulations discussed in this manuscript. All Ben~1 and Ben~2 simulations use fixed surface albedo (0.3), according to the THAI protocol \citep{Fauchez2020THAI}. More information on the planet's parameters used in THAI simulations are also provided in Table~\ref{tab:planet}. Note that \ce{CO2} condensation -- although it is expected for Ben~2 simulations \citep{Turbet:2018aa} -- has been disabled as not all the four GCMs possess this parameterization. Note also that gravity wave parameterizations have been turned off.

\begin{deluxetable*}{lll}
\tablecaption{Resolution used in the models.\label{tab:resolution}}
\tabletypesize{\scriptsize}
\tablehead{
\colhead{GCM} & \colhead{Number of grid points in the horizontal} & \colhead{Number of vertical levels (top layer)} }
\startdata
ExoCAM  & 72~$\times$~46 & 51 (1~Pa) \\
\midrule
\addlinespace[0.5cm]
LMD-G   & 72~$\times$~46 & 40 (4~Pa) \\
\midrule
\addlinespace[0.5cm]
ROCKE-3D & 72~$\times$~46 & 40 (10~Pa) \\
\midrule
\addlinespace[0.5cm]
UM & 144~$\times$~90 & 41 and 38 (4 and 13~Pa) \\
\enddata
\end{deluxetable*}

\begin{longrotatetable}
\begin{deluxetable*}{lllll}
\tablecaption{Dry physics parameterizations in the THAI GCMs.\label{tab:models}}
\tabletypesize{\scriptsize}
\tablehead{
\colhead{GCM} & \colhead{Radiative Transfer} & \colhead{Dynamical Core} & \colhead{Sub-grid dynamics} & \colhead{Land module}
}
\startdata
\addlinespace[0.2cm]
ExoCAM  & Correlated-$k$ & Finite-volume & 1\textsuperscript{st}-order closure & 15 ground layers \\
 & RT timestep of 1h  & Dynamical timestep of 56s & with explicit non-local flux of heat & total depth~=~35m \\
 & 0.2--1000$\mu$m  & Arakawa~C grid  & roughness~=~10$^{-2}$~m & Heat capacity~=~2.1~$\times$~10$^6$~J~m$^{-3}$~K$^{-1}$ \\
 & 17~$\times$~23 bands (SW~$\times$~LW)  & 4th-order divergence damping  & & Thermal inertia~=~4~$\times$~10$^3$~J~m$^{-2}$~K$^{-1}$~s$^{-1/2}$ \\
 & HITRAN~2004  & with Laplacian sponge & & \\
 & CO$_2$ self broadening~=~1.3$\times$~that of N$_2$ & 2nd-order velocity-component  & & \\
 & MT$\_$CKD~v2.5 CO$_2$ continuum ; N$_2$-N$_2$ CIA & damping for top layers & & \\
\addlinespace[0.2cm]
\midrule
\addlinespace[0.2cm]
LMD-G   & Correlated-$k$ & Finite-difference & 1.5-order (TKE) closure & 18 ground layers \\
 & RT timestep of 15min/7.5min (Ben1/Ben2) & Dynamical timestep of 90s/45s  & with dry convective adjustment &  total depth~=~20m \\
 & 0.3--1000$\mu$m  & (for Ben1/Ben2) & roughness~=~10$^{-2}$~m & Heat capacity~=~2~$\times$~10$^6$~J~m$^{-3}$~K$^{-1}$ \\
 & 36~$\times$~38 bands (SW~$\times$~LW) for Ben~1  & Arakawa~C grid & & Thermal inertia~=~2~$\times$~10$^3$~J~m$^{-2}$~K$^{-1}$~s$^{-1/2}$ \\
 & 36~$\times$~32 bands (SW~$\times$~LW) for Ben~2  &  & & \\
 & HITRAN~2012  & scale-selective horizontal dissipation & & \\
 & CO$_2$ self broadening~=~that of N$_2$ & based on an n~time iterated Laplacian & & \\
 & CO$_2$ sub-lorentzian far wings & & & \\
 & CO$_2$-CO$_2$ dimer~+~CIA ; N$_2$-N$_2$ CIA & & & \\
\addlinespace[0.2cm]
\midrule
\addlinespace[0.2cm]
ROCKE-3D & Correlated-$k$ & Finite-difference & 1.5-order (TKE) closure & 6 ground layers\\
 & RT timestep of 1h & Dynamical timestep of 225s/450s & with explicit non-local flux of heat & total depth~=~3.5m \\
 & 0.2--10000$\mu$m  & (for Ben1/Ben2) & roughness~=4.1~$\times$~10$^{-2}$~m & Heat capacity~=~0.85~$\times$~10$^6$~J~m$^{-3}$~K$^{-1}$ \\
 & 21~$\times$~12 bands (SW~$\times$~LW) for Ben~1  & Arakawa~B grid & & Thermal inertia~=~1.2~$\times$~10$^3$~J~m$^{-2}$~K$^{-1}$~s$^{-1/2}$ \\
 & 42~$\times$~17 bands (SW~$\times$~LW) for Ben~2  &  & & \\
 & HITRAN~2012  & numerical sponge in the top~6 layers & & \\
 & CO$_2$ self broadening~=~that of N$_2$ & & & \\
 & CO$_2$ sub-lorentzian far wings & & & \\
 & CO$_2$-CO$_2$ CIA ; no N$_2$-N$_2$ CIA & & & \\
\addlinespace[0.2cm]
\midrule
\addlinespace[0.2cm]
UM & Correlated-$k$ & Finite-difference & 1\textsuperscript{st}-order closure & 1 ground layer \\
 & RT timestep of 1h & Dynamical timestep of 1200s/600s & with explicit non-local & total depth~=~1m \\
 & 0.2--10000$\mu$m  & (for Ben1/Ben2) & fluxes of heat and momentum & Heat capacity~=~2~$\times$~10$^6$~J~m$^{-3}$~K$^{-1}$ \\
 & 21~$\times$~12 bands (SW~$\times$~LW) for Ben~1  & Arakawa~C grid & roughness~=~10$^{-2}$~m & Thermal inertia is implicitly infinite \\
 & 42~$\times$~17 bands (SW~$\times$~LW) for Ben~2  &  & & \\
 & HITRAN~2012  & numerical damping of vertical wind & & \\
 & CO$_2$ self broadening~=~that of N$_2$ & with altitude and latitude & & \\
 & CO$_2$ sub-lorentzian far wings & & & \\
 & CO$_2$-CO$_2$ CIA ; no N$_2$-N$_2$ CIA & & & \\
\addlinespace[0.2cm]
\enddata
\end{deluxetable*}
\end{longrotatetable}

\subsection{Models}
As specified in the THAI protocol \citep{Fauchez2020THAI}, four GCMs were used in this study:
\begin{itemize}
    \item the Exoplanet Community Atmospheric Model (ExoCAM), a branch of the National Center for Atmospheric Research Community Earth System Model version 1.2.1.
    \item the Laboratoire de Météorologie Dynamique - Generic model (LMD-Generic a.k.a. LMD-G).
    \item the Resolving Orbital and Climate Keys of Earth and Extraterrestrial Environments with Dynamics (ROCKE-3D Planet\_1.0),
    \item the Met Office Unified Model (UM, science version GA7.0, code version 11.6).
\end{itemize}

In the following subsections, we provide a brief description of the model components related to the dynamics and dry physics (see Part~2 of the THAI intercomparison for details on the moist physics), with a focus on RT, large-scale dynamics (computed using so-called `dynamical cores') and sub-grid scale dynamics (dry convection and turbulent mixing). Table~\ref{tab:resolution} gives an overview of the dynamical cores, as well as  the horizontal and vertical resolutions used in the four 3D models for all THAI simulations, including the Hab~1 \& Hab~2 cases analyzed in \citet{Sergeev21_THAI}. Table~\ref{tab:models} summarizes the dry physics parameterizations used by the models.

\subsection{Radiative Transfer} \label{subsec:RT}

To take into account how the atmosphere, clouds and surface interact with the light emitted by the star and the planet, GCMs employ RT schemes.

\medskip

ExoCAM includes a two-stream correlated-$k$ distribution RT scheme with a code lineage that traces back to \citet{UrataToon2013a, UrataToon2013b} and \citet{Colaprete2003} with extensive modifications and reorganization occurring over time \citep{Wolf&Toon2013, Kopparapu2017, Wolf2022}.  This RT scheme, known as ExoRT, has been separated from the 3D model and is provided on GitHub\footnote{https://github.com/storyofthewolf/ExoRT} in order to provide an accessible 1D column model version that facilitates more rapid development and testing. There are now several different RT versions available with ExoRT for use as part of ExoCAM that permit the simulation of different atmospheres.  For the THAI simulations, we utilized the oldest and most published RT version, named \textit{n28archean} in the ExoRT repository. This RT version was originally designed and tested for Archean Earth-like conditions, for \ce{CO2} amounts up to no more than ~30\%, Earth-Like water vapor column amounts, and Sun-like stellar inputs \citep{Wolf&Toon2013}.  Later studies have found this version of the RT to overestimate the greenhouse effect from pure \ce{CO2} atmospheres \citep{Wolf2022}.  Note that later iterations of ExoRT have corrected the issue with RT as is discussed in \citet{Kopparapu2017} and \citet{Wolf2022}.  The \textit{n28archean} version features 28 spectral intervals that extend from 0.2 to \SI{1000}{\micro\m}, with 8 Gauss points per bin.  The Gauss point binning used originates from RRTMG \citep{Mlawer1997} and is weighted towards 1 and thus favors capturing the peak of the absorption lines in each band.  While the longwave and shortwave stream bandpasses are customizable to optimize computational expense versus completeness, here we used a brute force approach and calculated the longwave and shortwave streams over the full spectrum in each case.  (Note, for planets in the TRAPPIST-1 system, about 1\% of the stellar energy received lies longward of \SI{9}{\micro\m})  Molecular absorption by \ce{CO2} is included, with correlated-k distributions produced using the Atmospheric Environment Research Inc. line-by-line RT model \citep[LBLRTM,][]{Clough:2005} using the HITRAN 2004 spectroscopic database \citep{Rothman2005}, and Voigt line wings cut at \SI{25}{\per\cm} for both \ce{H2O} and \ce{CO2}. Water vapor and \ce{CO2} continuum absorption are included using MT\_CKD version 2.5, while assuming that \ce{CO2} self-broadened continuum is 1.3 times the foreign broadening continuum \citep{Kasting1984, Halevy2009}. Recent improvements in the RT module of ExoCAM \citep{Wolf2022} – subsequent to this THAI intercomparison project – have shown that this approach overestimates the greenhouse effect of \ce{CO2} compared to using a sub-Lorentzian \ce{CO2} profile \citep{Perrin:1989,Tran2011} with a 500~cm$^{-1}$ cutoff, in agreement with the calculations of \citet{Wordsworth:2010}. To treat overlapping gas absorption the amount weighted scheme of \citet{Mlawer1997} is used. Collision induced absorption (CIA) is included for \ce{N2}-\ce{N2} pairs. Rayleigh scattering is treated using the parameterization of \citet{VardavasCarver1984} including \ce{N2} and \ce{CO2}.

The LMD-G GCM includes a flexible RT historically originating from the NASA Ames RT scheme, as described in \citet{Wordsworth:2011}. The RT is performed here for variable gaseous atmospheric compositions made of varying concentrations of \ce{CO2} and \ce{N2}, using the correlated-k method. The \ce{CO2} line list was taken from HITRAN~2012 \citep{Rothman:2013} and used to compute high-resolution absorption spectra (for multiple temperatures, pressures and molecular mixing ratios) based on Voigt profiles (truncation at \SI{500}{\per\cm}, following \citealt{Wordsworth:2010}) adjusted using experimentally-based $\chi$-factors \citep{Perrin:1989}. These high-resolution spectra were then converted into correlated-k coefficients, using 16 non-regularly spaced grid points for the g-space integration (a.k.a. gauss points), where g is the cumulative distribution function of the absorption data for each band. The correlated-k coefficients were also built using between 32 and 38 spectral bands in the thermal infrared (from 2.3 to \SI{1000}{\micro\m}) and between 32 and 36 spectral bands in the visible domain (from 0.3 to \SI{5.1}{\micro\m}), depending on the atmospheric composition considered. Besides, several continuum absorptions (CIA, dimer, far wings) were added for \ce{N2} (\ce{N2}-\ce{N2} CIA from \citealt{Richard:2012}) and \ce{CO2} (\ce{CO2}-\ce{CO2} CIA and dimer absorption from \citealt{Gruszka:1997} and \citealt{Baranov:2004}). LMD-G RT uses a two-stream scheme \citep{Toon:1989} to take into account the scattering effects of the atmosphere (Rayleigh scattering) and the clouds (Mie scattering ; but note that this is only relevant for Hab~1 and Hab~2 simulations), using the method of \citet{Hansen:1974}.

\medskip

Both the UM and ROCKE-3D share a common RT module SOCRATES\footnote{\url{https://code.metoffice.gov.uk/trac/socrates}}, which is an open-source two-stream correlated-$k$ RT model provided by the Met Office \citep{Edwards96,Manners12}.  SOCRATES is highly flexible. \textit{Spectral files}, created for both longwave and shortwave streams separately, are tailored to specific atmospheric compositions and stellar spectra.  The spectral files are then input to the GCMs at run-time, containing all information necessary to define the RT problem, including the number of spectral intervals, the absorbing gases, continua, and CIAs, the number of Gauss points used for each gas, the designation of dominant and minor absorbing species in each interval, and cloud and Rayleigh scattering properties.  Thus, dramatic changes to the RT can be achieved by careful pre-processing of the spectral files, with no changes to the hard-code being required.  When run embedded in a GCM, SOCRATES uses the equivalent extinction method to treat multiple overlapping gas species \citep{Amundsen_2017}.  In the equivalent extinction method the dominant absorbing gas, as determined from the total transmissivity in a test column, is treated with a full k-distribution, while all minor absorbing gases are treated as grey.  The number of Gauss points used for the dominant gas in each spectral interval is determined based on a selected transmission error tolerance (usually ~0.001). For the Ben~1 simulations, spectral files were tailored to Earth-like atmospheric compositions, allowing for small amounts of CO$_2$ in N$_2$ dominated atmospheres. 
The spectral files used by ROCKE-3D and the UM are identical (details provided in Table~\ref{tab:models}). 
Gas absorption coefficients are derived from the HITRAN~2012 spectroscopic database \citep{Rothman:2013}.  For Ben~2 simulations, CO$_2$ absorption is treated following \citet{Wordsworth:2010} with sub-Lorenztian line shape \citep{Perrin:1989} out to \SI{500}{\per\cm} with \ce{CO2}-\ce{CO2} CIA overlaid.

\subsection{Dynamical Cores (and numerical damping)} \label{subsec:dynamical_core}
A key part of a GCM is its dynamical core, which solves a system of Navier-Stokes equations on a spherical grid to simulate movement of mass and energy in the atmosphere.
To solve these equations, three THAI GCMs (LMD-G, ROCKE-3D, the UM) use a finite differences approach, while ExoCAM is based on a finite volume approach; there are also important differences in the implementation and approximations used.

\medskip
ExoCAM uses a finite volume scheme \citep{lin&rood:1996}.  Modifications have been brought to its dynamical core to improve numerical stability in more strongly forced atmospheres. They consist in incrementally applying physics tendencies for temperature and wind speed evenly throughout the dynamical time step, rather than only at the beginning of it \citep{bardeen:2017}.
The number of points in the latitudinal and longitudinal directions is 72 and 46, respectively.
This gives the horizontal grid spacing of \ang{5}$\times$\ang{3.75}.
ExoCAM uses 51 levels in the vertical, stretching from the surface to \SI{1}{\hecto\pascal} in the atmosphere. Regarding numerical damping, a 4th-order divergence damping with Laplacian sponge is applied. A 2nd-order velocity-component damping is also used for top layers. There is therefore an implicit numerical sponge layer produced by the degradation of the order of the numerical scheme.

The LMD-G's GCM dynamical core solves the primitive equations of meteorology using an enstrophy and total angular momentum conserving finite difference dynamical core on an Arakawa-C grid (for more details, see \citealt{Forget:1999} and \citealt{Hourdin:2006}). A filter is applied at high latitudes to satisfy the Courant–Friedrichs–Lewy numerical stability criterion. For the THAI simulations, LMD-G uses a uniform resolution of \ang{5} in longitude and \ang{3.75} in latitude. The dynamical time-step is \SI{45}{\s}. The model uses hybrid vertical coordinates (terrain-following coordinate system in the lower atmosphere; fixed pressure levels in the upper atmosphere). For the THAI simulations, we used 40~vertical layers, with the lowest mid-layer
level at \SI{5}{\m} and the top mid-layer level at \SI{\approx 4}{\pascal}. Non-linear interactions between explicitly resolved scales and subgrid-scale processes (which are necessary to ensure numerical stability) are parameterized by applying a scale-selective horizontal dissipation operator based on an $n$~time iterated Laplacian as described in \citet{Forget:1999}, Section~3.

\medskip
The ROCKE-3D dynamical core utilizes finite differences where atmospheric velocity points are located on the Arakawa-B grid \citep{Arakawa1972,Hansen1983}. The atmospheric vertical layers use a sigma coordinate system from the surface to 150~hPa. From there constant pressure layers are used to the TOA ($\approx$~0.1~hPa). For the THAI experiments the model was run with 72 $\times$ 46 grid points with each grid point being \ang{5} in longitude by \ang{4} in latitude except at the poles where the latitude coordinate is only \ang{2} in extent. The smaller polar grid boxes are necessary to maintain second-order accuracy of airmass fluxes, the pressure gradient force, and momentum advection terms
\citep{Schmidt2006}. The dynamical time-step used in the Ben~1 simulation was \SI{225}{\s} and \SI{450}{\s} in Ben~2. 
Both Hab~1 and Hab~2 simulations use a dynamical time-step of \SI{450}{\s}.
For all 4 simulations the parameterized physics uses a
time-step of $\approx$ 30 minutes (\SI{1756.8}{\s} to be exact), and less in some submodules, while the SOCRATES radiation was called every two physics time-steps ($\sim$ 1 hour).

\medskip
The UM employs the ENDGAME dynamical core with a semi-implicit semi-Lagrangian formulation to solve the non-hydrostatic fully compressible deep-atmosphere equations of motion \citep{Wood2014}.
Model fields are discretized horizontally onto a regular longitude-latitude Arakawa-C grid, while in the vertical a terrain-following hybrid height coordinate with a Charney-Phillips staggering is used.
THAI experiments are run with $144\times 90$ grid points in the horizontal, giving a grid spacing of \ang{2.5} in longitude and \ang{2} in latitude.
In the vertical, 41 levels between the surface and a fixed model lid, located at \SI{\approx 63}{\km} height, are used for the Ben~1 experiment.
For Ben2, the number of levels is 38 and the model lid is at \SI{\approx 40}{\km} to improve the model stability.
The primary dry prognostic variables in the UM are the three wind components, virtual dry potential temperature, Exner pressure\footnote{Exner pressure is defined as $\Pi=(p/p_0)^{R_d/c_p}$, where $p$ is the air pressure, $p_0$ is the reference pressure, $R_d$ is the specific gas constant, and $c_p$ is the specific heat capacity at constant pressure.} and density.
Increments to these variables are obtained on each dynamical time step (\SI{1200}{\s} in Ben~1, \SI{600}{\s} in Ben~2), within which processes are split into an outer loop.
The semi-Lagrangian departure points are obtained within the outer loop using the latest estimates for the wind components.
Fields are then interpolated to the updated departure points.
Within the inner loop, a linear Helmholtz problem is solved to calculate the pressure increment.
Estimates for all variables are then obtained from the pressure increment via a back-substitution process \citep[for a schematic of the time-stepping algorithm and more details, see][]{Walters17}.
There is no explicit diffusion or dissipation in the model.
At the TOA, a sponge layer is used to suppress numerical instabilities by explicitly damping vertical velocity using a $sin^2$ function.
The damping is applied in the upper half of the model domain above at equator, dropping in altitude with latitude, eventually to surface at the poles.

\subsection{Sub-grid scale mixing} \label{subsec:dry_convection}
To take into account how the atmosphere acts on scales too small to be resolved by the dynamical cores, GCMs employ subgrid-scale parameterizations.
For a dry simulation, they include parameterizations of turbulence and dry convection, which are usually handled by the boundary-layer or convection schemes, or a combination of both.

\medskip

The ExoCAM parameterization package \citep[CAM4,][]{Neale2010} can be described by a sequence of four components: the moist precipitation processes, the cloud and radiation, the surface model and the turbulent mixing.
In the Ben1 \& Ben 2 cases, only the shallow part of convection is active to by ensuring that the lapse rate is stable and stratified.
The surface model include surface fluxes obtained from land, ocean and sea ice models, or computed from specific conditions.
These surface fluxes provide lower flux boundary conditions for the turbulent mixing which is comprised of the planetary boundary layer (PBL) parameterization, vertical diffusion.

In LMD-G, subgrid-scale dynamical processes including turbulent mixing and convection are parameterized as in \citet{Forget:1999}. In practice, the boundary layer dynamics are accounted for by the \citet{Mellor:1982} unstationary 2.5-level closure scheme (using the implementation of \citealt{Galperin:1988}) plus a convective adjustment which rapidly mixes the atmosphere in the case of unstable temperature profiles. In that case, the thermal profile is brought back to the dry or wet adiabat (in the case where water vapour condenses, which is only relevant for Hab~1 and Hab~2 simulations described in THAI part~2). In the Ben~1 and Ben~2 (dry) simulations, only the former case applies. Turbulence and convection mix energy (potential temperature), momentum (wind), and tracers (gases and aerosols) as described in \citet{Forget:1999}, Section~6.1. For the THAI simulations, a standard roughness coefficient $z_0=\SI{e-2}{\m}$ is used for both dry (Ben~1 and Ben~2) and ocean (Hab~1 and Hab~2, see THAI part~2) surfaces for simplicity. 
\medskip

In ROCKE-3D, the level 2 of the \citet{Cheng2002} model is used in the free troposphere (above the PBL - which is determined dynamically). This scheme has the ability to produce fluxes when turbulence is weak, at large Richardson numbers (large buoyancy to shear forcing ratios).
From the mid-point of the first atmospheric model layer to the surface there are 8 sublayers used to compute surface turbulent fluxes with the level 2.5 model from \citet{Cheng2002}.  
In between the two it depends upon the stability of the PBL.
In the stable case: we use level 2 of the \citet{Cheng2002} model, which increases the critical Richardson number (from \citealt{Mellor:1982} model’s value 0.2 to 1), in agreement with several data sets, and with the potential to increase the PBL height (the latter is Richardson number related).
In the unstable case: for diffusivities, we use the non-local model of \citet{Holtslag1993} with counter-gradient fluxes, and for the turbulent kinetic energy (TKE) we use the parameterization based on the large eddy simulation (LES) by \citet{Moeng1994}. 
For details see \citet{Schmidt2006} and references therein.

\medskip
To represent mixing of adiabatically conserved heat, momentum and tracers throughout the troposphere, including the PBL, the UM uses a first-order turbulence closure of \citet{Lock00} with modifications described in \citet{Lock01} and \citet{Brown08}.
The scheme treats unstable and stable boundary layers differently.
For unstable layers, two separate profiles of diffusion coefficients, one for the surface sources of turbulence and one for upper atmospheric sources (e.g. radiative cooling).
Initially diagnosed by an adiabatic ascent, the extent of these profiles is adjusted so that the magnitude of the buoyancy consumption of turbulent kinetic energy is limited to a specified fraction of buoyancy production.
In moist aquaplanet simulations (see Part II), this permits the cloud layer to decouple from the surface.
Additional non-local fluxes of heat and momentum can generate more vertically uniform potential temperature and wind profiles in unstable (convective) boundary layers.
The scheme is tightly coupled to the convection scheme (see Part II), but the latter is inactive in dry cloudless simulations of the Ben~1 \& 2 cases.
For stable boundary layers and the free troposphere, diffusion coefficients are calculated using the parameterization of \citet{Smith90}, based on a local Richardson number.
The mixing can be also enhanced if the non-local coefficients, calculated as described above, are larger than the local ones, which can happen in the neutral boundary layer.
The stability dependence is given by the ``MES-tail'' function within the stable boundary layer and by the ``sharp'' function at \SI{200}{\m} and above \citep{Brown08}.
Given the resulting coefficient, the diffusion equation is solved implicitly, and the kinetic energy dissipated through the turbulent shear stresses is converted to a local heating term.

\section{Results} \label{sec:results}

\begin{deluxetable}{lccc}
\renewcommand{\arraystretch}{1.25}
\tablecaption{Global mean surface temperature (T\textsubscript{s}, \si{\K}) global mean top-of-atmosphere outgoing longwave radiation (F\textsubscript{IR}, \si{\watt\per\square\meter}), planetary albedo ($\alpha_\text{p}$) in Ben~1 and Ben~2 simulations.\label{tab:glob_diag}}
\tablewidth{0pt}
\tablehead{
GCM & \multicolumn{3}{c}{Ben~1} \\
{} & T\textsubscript{s} & F\textsubscript{IR} & $\alpha_\text{p}$}
\startdata
ExoCAM  & 220 & 160 & 0.29 \\
LMDG    & 217 & 160 & 0.30 \\
ROCKE3D & 216 & 164 & 0.27 \\
UM      & 213 & 162 & 0.28 \\
\midrule
{} & \multicolumn{3}{c}{Ben~2} \\
ExoCAM  & 245 & 179 & 0.21 \\
LMDG    & 242 & 174 & 0.22 \\
ROCKE3D & 242 & 184 & 0.18 \\
UM      & 239 & 182 & 0.19 \\
\enddata
\end{deluxetable}

We present and compare results from the four GCMs onboard the THAI intercomparison, for the dry Ben~1 and Ben~2 simulations of TRAPPIST-1e. We first discuss radiation fluxes, then convection and turbulence, and eventually large-scale dynamics.

\subsection{Radiative budget and fluxes} \label{sec:results_RB}

\begin{figure*}
\centering 
\subfloat[Results for the Ben~1 simulations]{
\label{fig:sub-first_ASR}
\includegraphics[width=0.8\linewidth]{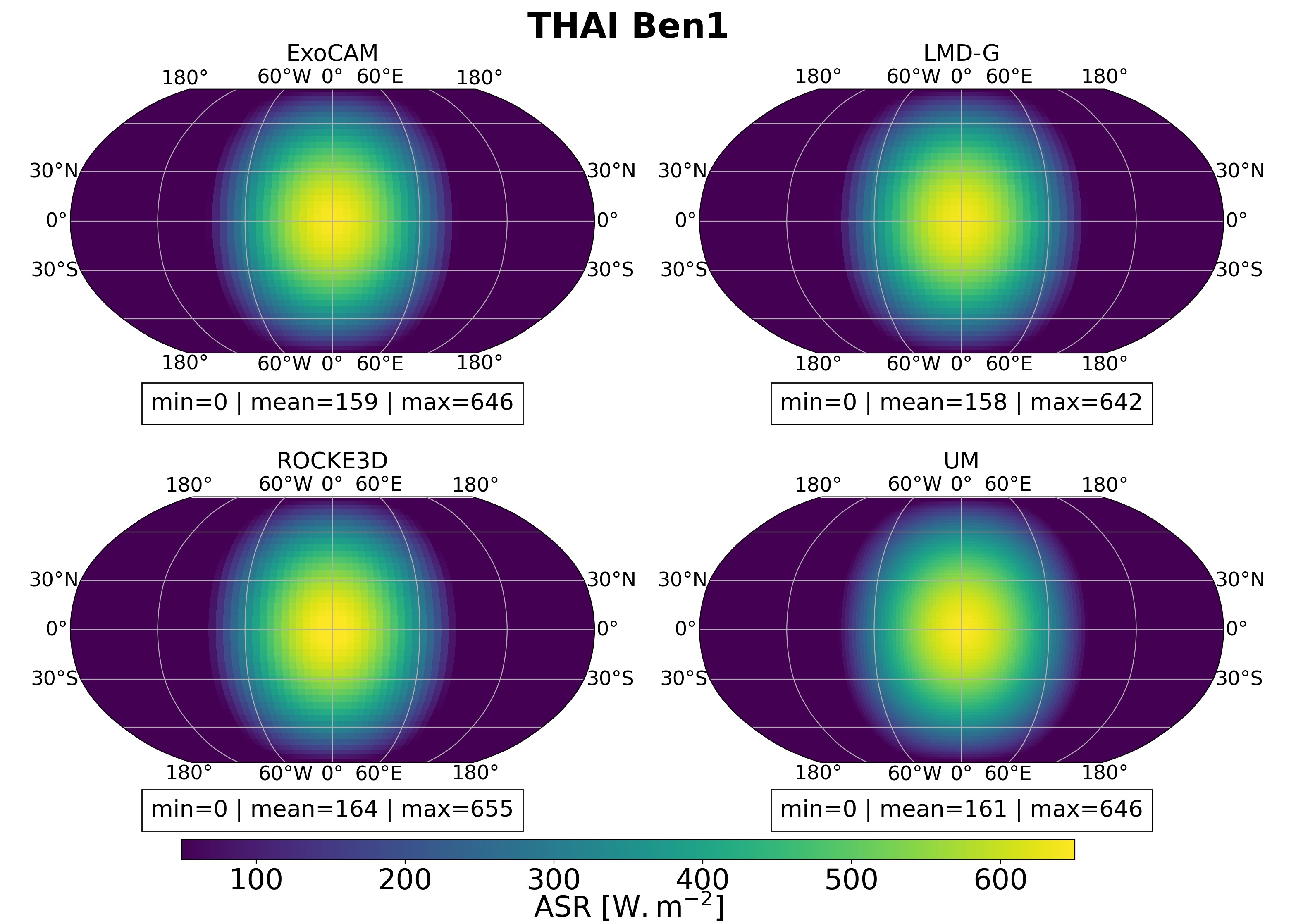}}
\\
\subfloat[Results for the Ben~2 simulations]{
\label{fig:sub-second_ASR}
\includegraphics[width=0.8\linewidth]{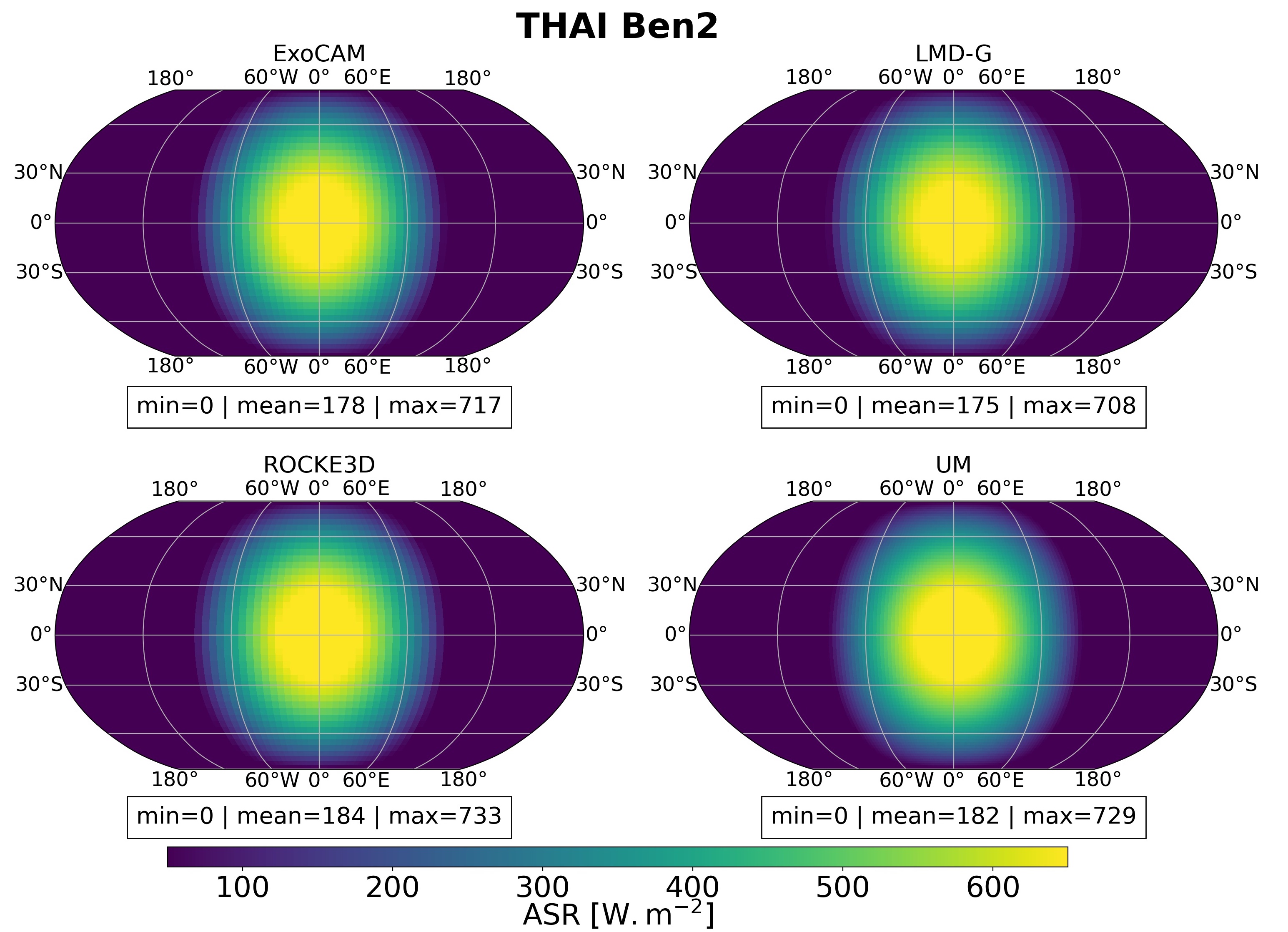}}
\caption{Absorbed Stellar Radiation (ASR) maps for the Ben~1 and Ben~2 THAI simulations, for the four GCMs. The temporal average minimum, mean and maximum values are also shown below each map.
\label{fig:fig_ASR}}
\end{figure*}

In order to compare the radiative budgets of the GCMs, we first start with the shortwave fluxes (which should not be very sensitive to the thermal structure of the atmosphere in the Ben~1 and Ben~2 dry cases) and then compare the thermal, longwave fluxes.
Fig.~\ref{fig:fig_ASR} shows the absorbed stellar radiation (ASR) horizontal map for the Ben~1 (left) \& Ben~2 (right) cases. ASR is defined here as the total flux from the star (TRAPPIST-1) that is absorbed by the planetary surface and atmosphere. In both the Ben~1 and Ben~2 cases, the mean and maximum values agree between the models within a few \si{\watt\per\m\squared}. The minimum values are equal to \SI{0}{\watt\per\m\squared} for all GCMs as the nightside is not illuminated. The highest ASR values are for ROCKE-3D followed by the UM, ExoCAM and LMD-G GCMs. 

Since the surface albedo is fixed at 0.3 in the THAI Ben~1 and Ben~2 simulations (as defined in the THAI protocol of \citealt{Fauchez2020THAI}) and the stellar spectral flux is the same for the four GCMs, differences in absorbed flux may arise from Rayleigh scattering but more likely from molecular absorption (from \ce{CO2 and marginally from N$_2$-N$_2$ CIA}, in the Ben~1 and 2 simulations) because the spectrum of the star TRAPPIST-1 is very strongly shifted toward infrared wavelengths (see e.g., \citealt{TurbetSelsis2021}, Fig.~1) where \ce{CO2} can absorb (and where Rayleigh scattering has a very limited role). For the Ben~1 simulations, the mean Bond albedo is equal to 0.27/0.28/0.29/0.3 for ROCKE-3D/UM/ExoCAM/LMD-G, respectively (see Table~\ref{tab:glob_diag}). These values are very close to the surface albedo, indicating little influence from Rayleigh scattering and \ce{CO2} absorption (only \SI{\approx 400}{ppm} of \ce{CO2} in the Ben~1 simulations). For the Ben~2 simulations, the mean Bond albedo is equal to 0.18/0.19/0.21/0.22 for ROCKE-3D/UM/ExoCAM/LMD-G, respectively (see Table~\ref{tab:glob_diag}). These values are significantly lower than that of the Ben~1 simulations because here the atmosphere is essentially composed of 1~bar of \ce{CO2} (about 2500$\times$ the amount in Ben~1) that can efficiently absorb the incoming stellar radiation in the near-infrared. 

\begin{figure*}
\centering 
\includegraphics[width=0.8\linewidth]{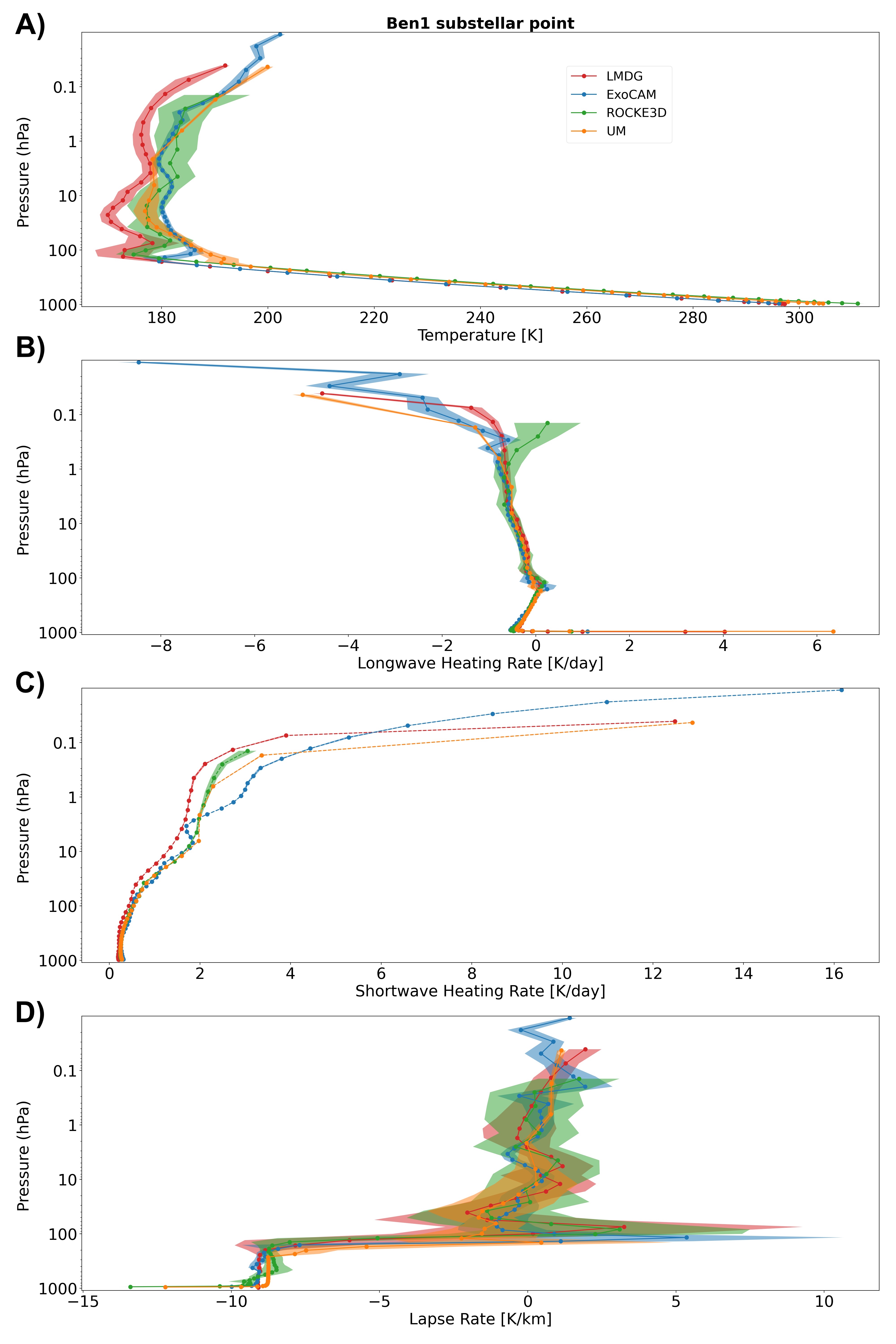}
\caption{Vertical profiles at the substellar point for the Ben~1 simulations of (A) the temperature, (B) the net longwave heating rate (i.e. the heating rate of each layer resulting from a balance between thermal infrared emission of the layer and the absorption of thermal infrared emission of the other layers and the surface), (C) the shortwave heating rate (i.e. the heating rate of each layer produced by the absorption of the stellar flux), (D) the lapse rate (i.e. the vertical temperature gradient). Time averaged values are represented by thick lines while the $1-\sigma$ deviations (over the 10~orbits) are represented by shades.
\label{fig:fig_substellar_Ben1}}
\end{figure*}

\begin{figure*}
\centering 
\includegraphics[width=0.8\linewidth]{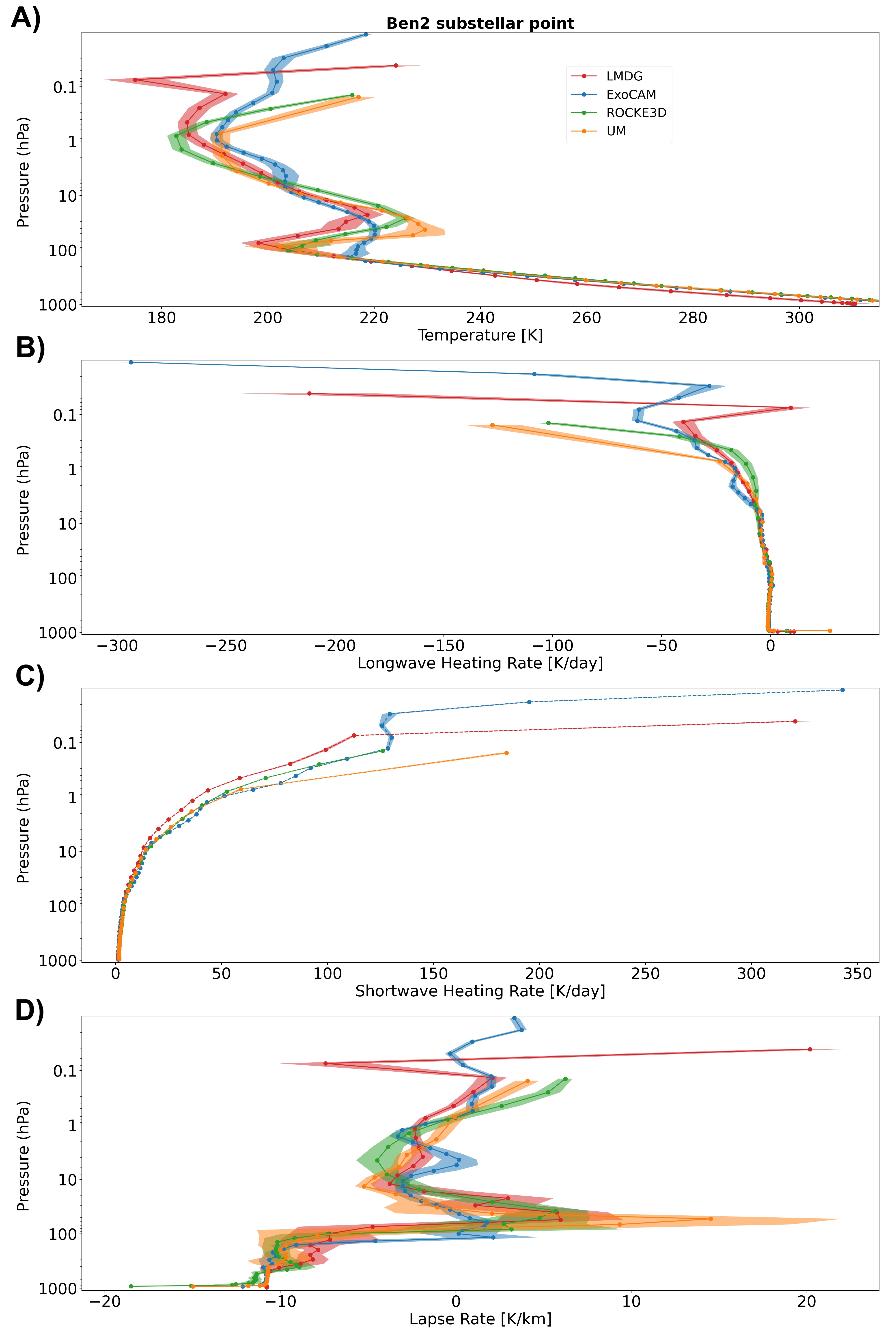}
\caption{Same as Fig.~\ref{fig:fig_substellar_Ben1} for the Ben~2 simulations.
\label{fig:fig_substellar_Ben2}}
\end{figure*}

The differences between the outputs obtained with our set of GCMs are very small and comparatively similar between the Ben~1 and Ben~2 cases. Similar results are also observed in the shortwave heating rate vertical profiles (see Fig.~\ref{fig:fig_substellar_Ben1}C and Fig.~\ref{fig:fig_substellar_Ben2}C), which display similar trends from the surface up to \SI{\approx 1}{\hecto\pascal}. The Ben~2 simulations result in a much stronger shortwave heating rate due to direct absorption of stellar light by \ce{CO2}. Both the Ben~1 and Ben~2 cases exhibit shortwave heating rates that increase as altitude increases (or pressure decreases). This is because CO$_2$ absorption saturates as we go deeper into the atmosphere (because the upper layers absorb all photons in the core of near-infrared CO$_2$ band lines).

Most of the differences in the RT likely come from differences in the parameterization of \ce{CO2} absorption (\ce{CO2} is the main absorbing species in Ben~1 and Ben~2 simulations). This is supported in particular by the fact that the two models --- ROCKE-3D and UM --- that share the same RT scheme, namely SOCRATES (as well as the same spectral files) display very similar shortwave heating rates from the surface up to \SI{\approx 1}{\hecto\pascal} (see orange and green lines in Fig.~\ref{fig:fig_substellar_Ben1}C and Fig.~\ref{fig:fig_substellar_Ben2}C). The two models diverge mainly for the two uppermost levels of UM (especially the last altitude level), presumably due to different RT boundary conditions between UM and ROCKE-3D.
At the pressures where the differences between shortwave heating rates are the most significant between models, CO$_2$ absorption is in principle dominated by the cores of CO$_2$ band lines (and not the far wings or CIA/dimer absorptions caused by molecular collisions). Note that the shortwave heating rate variability within individual models (identified by shades in Fig.~\ref{fig:fig_substellar_Ben1}C and Fig.~\ref{fig:fig_substellar_Ben2}C) is small, indicating that the near-infrared CO$_2$ absorption weakly depends on atmospheric temperature. A more detailed comparison of the correlated-k tables used by the models is required to identify the origins of the variations in the high-altitude shortwave heating rates.

\begin{figure*}
\centering 
\subfloat[Results for the Ben~1 simulations]{
\label{fig:sub-first_OLR}
\includegraphics[width=0.8\linewidth]{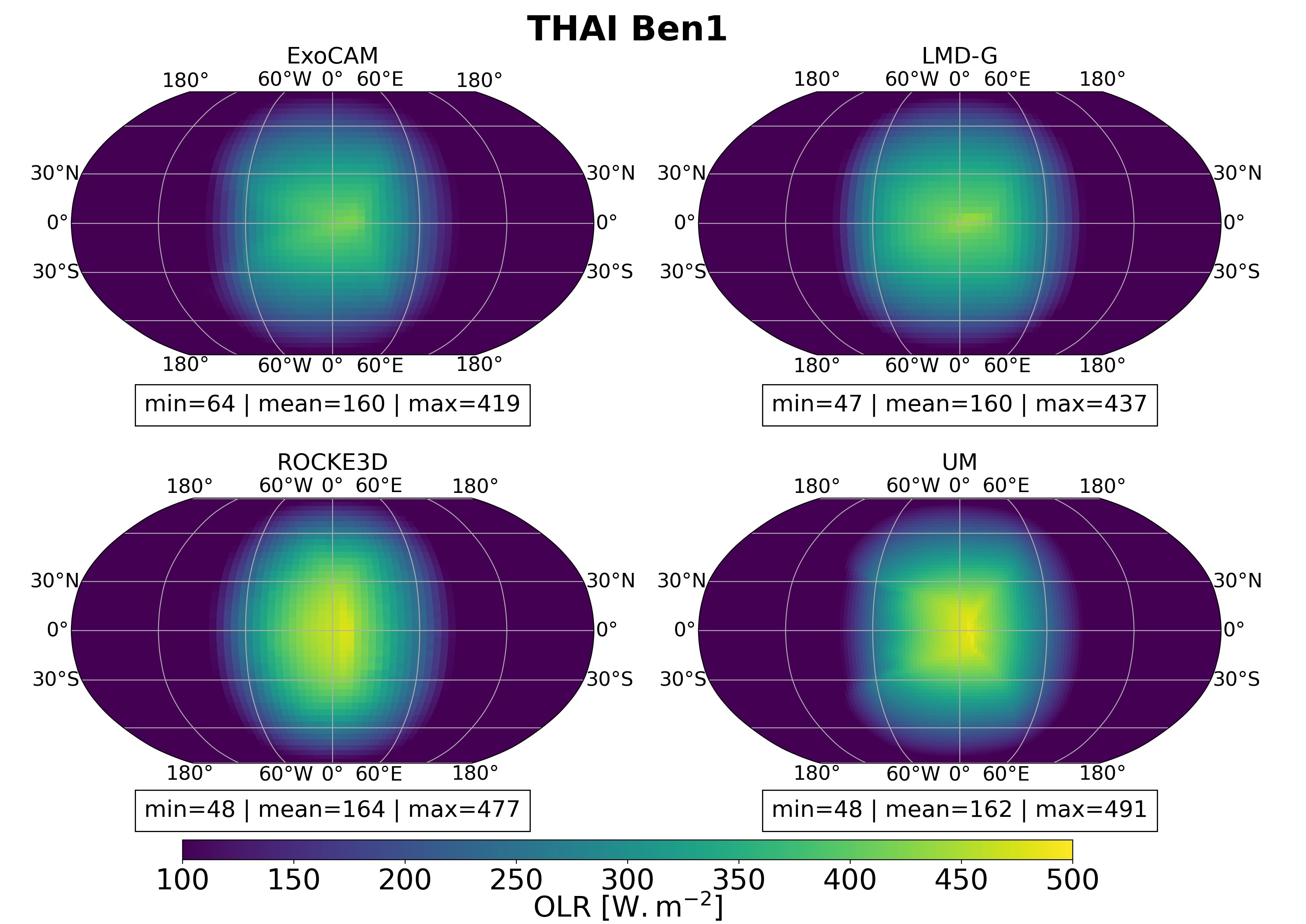}}
\\
\subfloat[Results for the Ben~2 simulations]{
\label{fig:sub-second_OLR}
\includegraphics[width=0.8\linewidth]{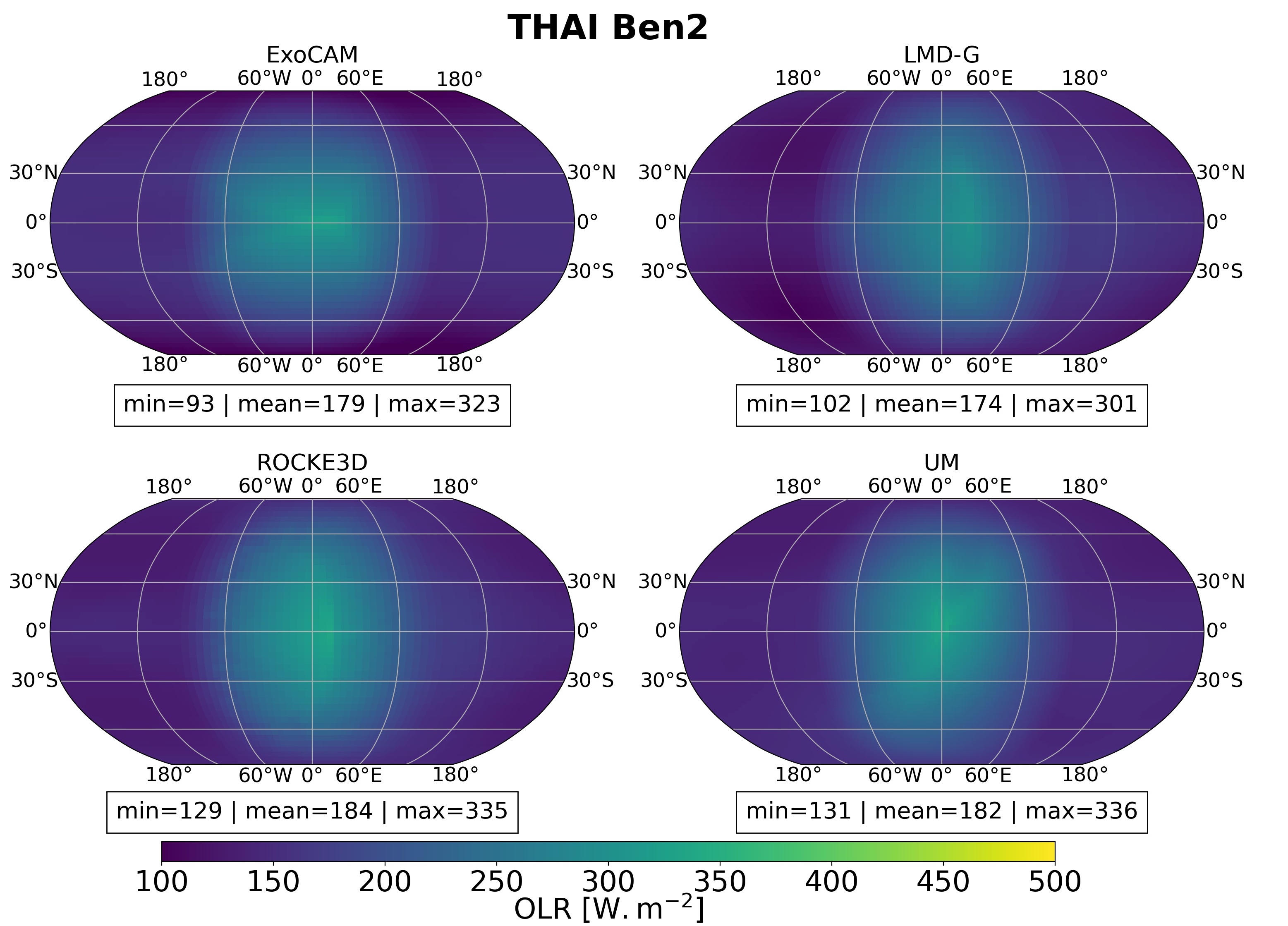}}
\caption{Same as Fig~\ref{fig:fig_ASR} but for the Outgoing Longwave Radiation (OLR).
\label{fig:fig_OLR}}
\end{figure*}

\begin{figure*}
\centering 
\subfloat[Results for the Ben~1 simulations]{
\label{fig:sub-first_tsurf}
\includegraphics[width=0.8\linewidth]{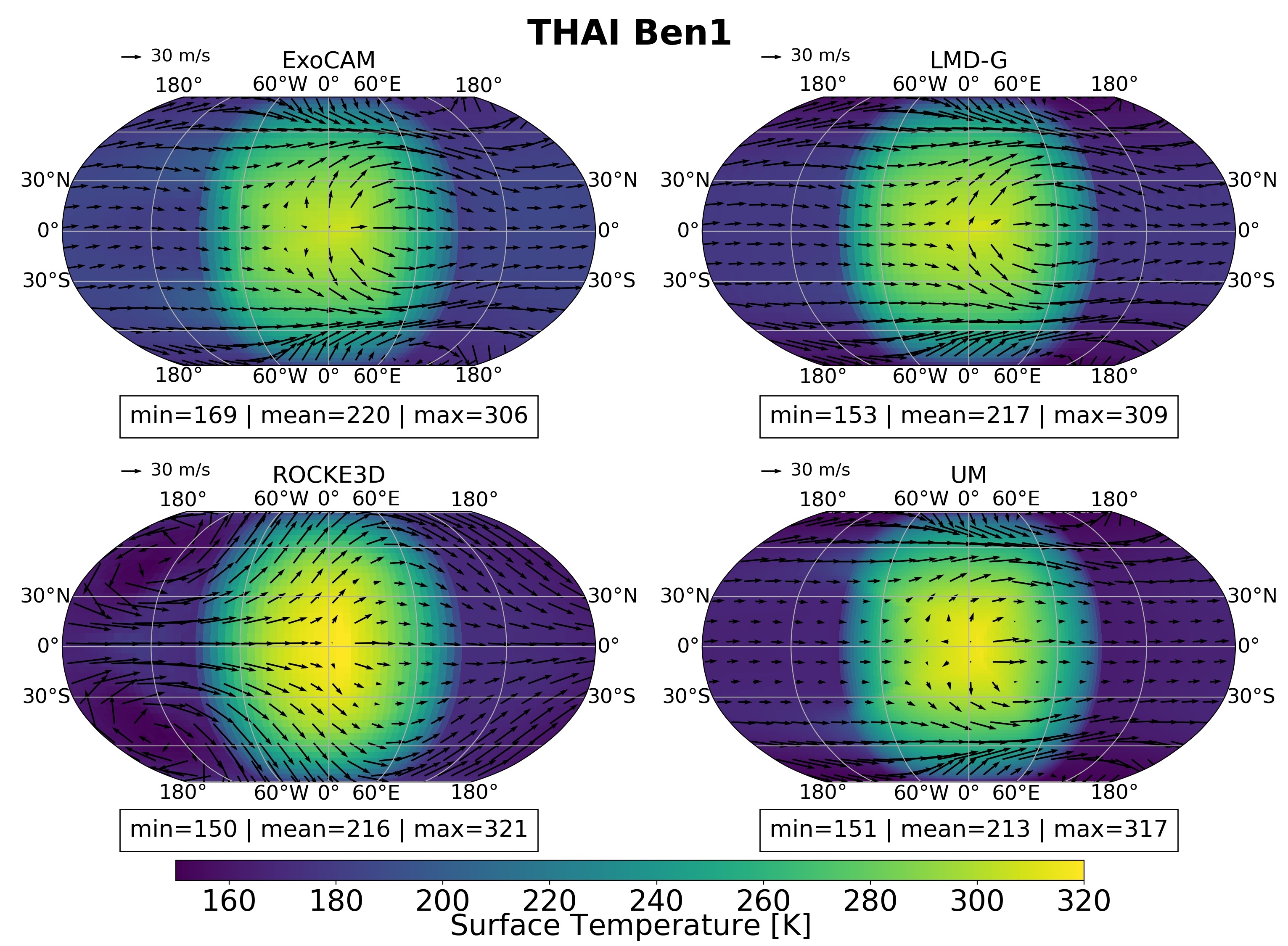}}
\\
\subfloat[Results for the Ben~2 simulations]{
\label{fig:sub-second_tsurf}
\includegraphics[width=0.8\linewidth]{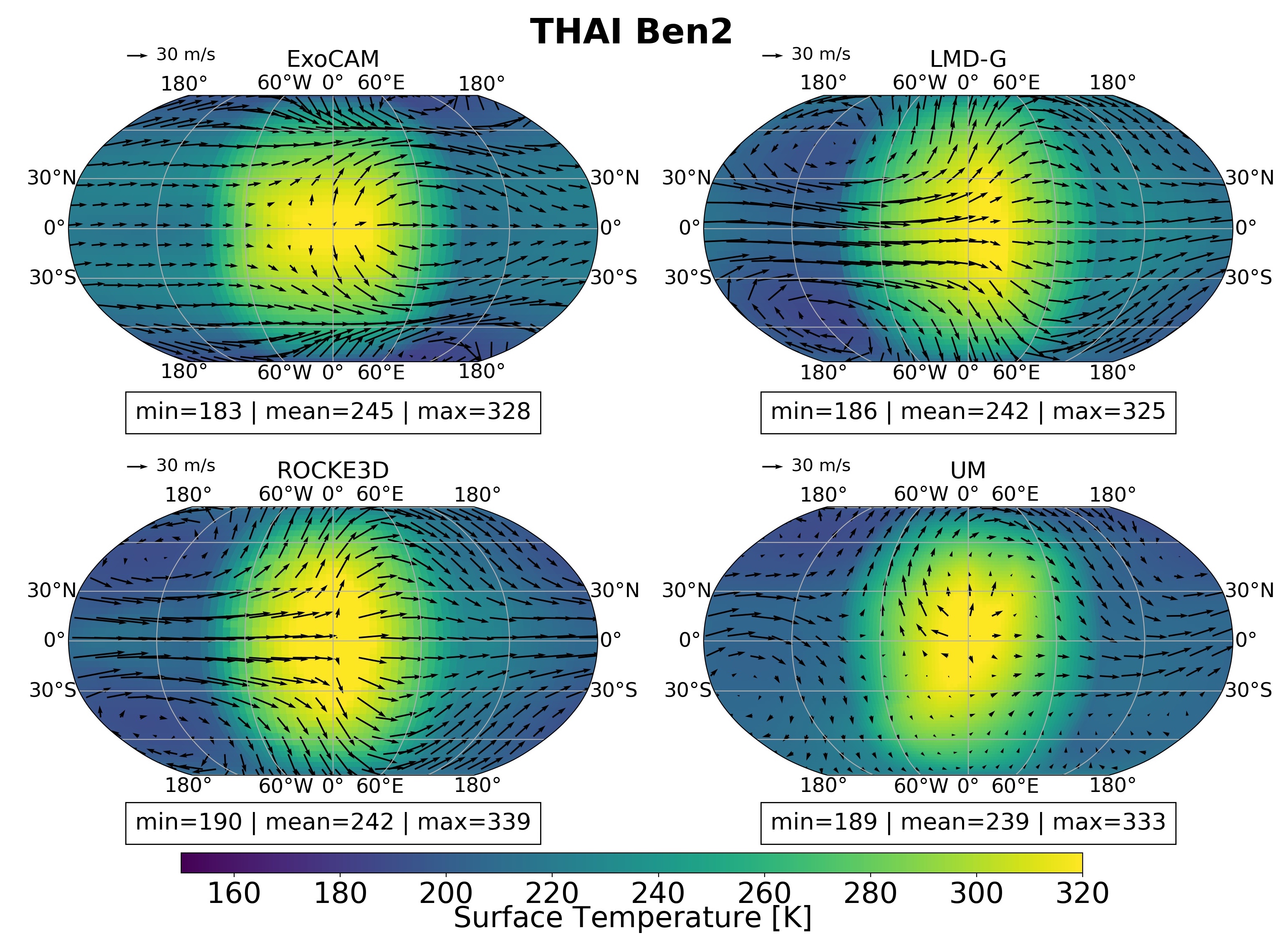}}
\caption{Surface temperature (shading, \si{\K}) maps on which we superimposed horizontal wind vectors in the upper troposphere (\SI{\approx 250}{\hecto\pascal})
\label{fig:fig_tsurf}.}
\end{figure*}

Fig.~\ref{fig:fig_OLR} is similar to Fig.~\ref{fig:fig_ASR} but for the TOA outgoing longwave radiation (OLR).
TOA OLR is the total amount of infrared radiation that is emitted at the top of the model.
For the Ben~1 case, the inter-model agreement is quite good, with LMD-G and ExoCAM showing exactly the same mean value of \SI{160}{\watt\per\m\squared} while the UM and ROCKE-3D have higher values by only 2 and \SI{4}{\watt\per\m\squared}, respectively (see Table~\ref{tab:glob_diag}).
The agreement is also good for the Ben~2 case (see Table~\ref{tab:glob_diag}), but the difference slightly increases up to \SI{10}{\watt\per\m\squared} between the mean values of LMD-G and ROCKE-3D, and up to \SI{35}{\watt\per\m\squared} between the maximum values of LMD-G and the UM (slightly eastward of the substellar point).
The global-mean values of OLR match those for ASR, confirming that the models are close to the radiative balance at the TOA.
The OLR maps are controlled by the temperature of the surface (Fig.~\ref{fig:fig_tsurf}) and the atmosphere, both of which are heated and cooled by the global heat redistribution.
Broadly, the OLR peak is located at or a few degrees to the east of the substellar point, which is where the surface absorbs most of the instellation and thus is the warmest. Weakest OLR regions are also co-located with coldest surface temperature regions.

At the substellar point, the tropospheric (from the surface up to 100~hPa) temperature structure is very similar across all models (in both Ben~1 and Ben~2 simulations), as seen in Fig.~\ref{fig:fig_substellar_Ben1}A and Fig.~\ref{fig:fig_substellar_Ben2}A, respectively. Under these conditions, the substellar longwave heating rate profiles in the troposphere (see Fig.~\ref{fig:fig_substellar_Ben1}B and Fig.~\ref{fig:fig_substellar_Ben2}B) agree well with each other, in both the Ben~1 and Ben~2 simulations. The longwave heating rates are net quantities: in each atmospheric layer they result from its thermal emission as well as the absorption of thermal radiation of the other layers and the surface. The longwave heating is relatively small near the surface for all models and for both the Ben~1 and Ben~2 simulations; while in the upper atmosphere strong longwave cooling dominates due to the effective infrared emission by CO$_2$, particularly prominent for the Ben~2 simulations (Fig.~\ref{fig:fig_substellar_Ben2}B). In the upper atmosphere (i.e. the radiative part), net longwave heating rate profiles are strongly anti-correlated with temperature profiles (in other words, longwave cooling is strongly correlated with temperature) because the temperature controls the emission of the layers.

In summary, the RT calculations, shown here for the substellar point, across all GCMs produce very similar results (with variations of radiative heating rates between the four models of always less than \SI{5}{\percent}, from the surface up to \SI{\approx 5}{\hecto\pascal}), except in the uppermost part of the atmosphere. Vertical variations of the net longwave heating rates are strongly linked to the temperature structure. The longwave heating varies horizontally across the planet too, which is a function of the temperature structure discussed in the next sections.

\subsection{Turbulence and Convection} \label{sec:results_TC}

In order to compare the behaviour of our GCMs for the convective boundary layer, we focus on the substellar point, where direct stellar heating of the surface drives intense convection (evident e.g. in the vertical winds pattern in Figs.~\ref{fig:fig_substellar_Hdecomposition_Ben1} and \ref{fig:fig_substellar_Hdecomposition_Ben2}).

As was mentioned in the previous section, the GCMs agree well on the temperature profile at the substellar point. Subtle differences are revealed by looking at the lapse rate (i.e. the vertical temperature gradient), shown in Fig.~\ref{fig:fig_substellar_Ben1}D and \ref{fig:fig_substellar_Ben2}D. In the troposphere, the lapse rates are about -9 and \SI{-11}{\K\per\km} on average for the Ben~1 and Ben~2 cases, respectively. To first order, the difference between the Ben~1 and Ben~2 simulations comes from the variation of the atmospheric heat capacity between the cases (about 1040 and 850~J~kg$^{-1}$~K$^{-1}$ for Ben~1 and Ben~2, respectively).
Note that because there are no changes in the atmospheric composition in our simulations, all four GCMs have the heat capacity set to a fixed parameter in each of the cases.

ExoCAM, ROCKE-3D and the UM show super-adiabatic behaviour near the surface, while LMD-G --- whose temperature profile is driven by a dry convective adjustment scheme --- follows closely the $-g/c_p$ lapse rate. 
Dry convection is triggered by the instability induced by the presence of a super-adiabatic layer near the surface. In the case of LMD-G, the convective adjustment instantly removes the super-adiabatic layer, forcing it to follow the $-g/c_p$ lapse rate. In the other three models, the super-adiabatic layer persists due to the (non-zero) timescale needed for the hot near-surface air parcels to rise.

The models all have a troposphere that reaches about 100~hPa, for both Ben~1 and Ben~2 simulations, in line with the theoretical prediction of \citet{Robinson:2014}. However, there are notable differences between models, in particular regarding the temperature of the tropopause. The inter-model variability is about 15~K for both Ben~1 and Ben~2 simulations (see Fig.~\ref{fig:fig_substellar_Ben1}A and \ref{fig:fig_substellar_Ben2}A). LMD-G consistently produces the coldest tropopause, as in the Hab~1 and Hab~2 aquaplanet simulations presented in THAI part 2 \citep{Sergeev21_THAI}. This is most likely due to the fact that LMD-G also has the weakest stratospheric shortwave heating rates in both Ben~1 and Ben~2 simulations, (see Fig.~\ref{fig:fig_substellar_Ben1}C and \ref{fig:fig_substellar_Ben2}C).

Despite these differences, we note that the convective layer of the atmosphere is the part that produces the smallest temperature variations among the models (as well as the smallest internal variability in the models, represented by the shades in Fig.~\ref{fig:fig_substellar_Ben1}A and \ref{fig:fig_substellar_Ben2}A). This is a good marker that stellar energy input is consistent across the four GCMs.

\subsection{Large Scale Dynamics} \label{sec:results_LSD}

To understand and interpret the horizontal variations in OLR and temperature between the GCMs, and more generally the climate and circulation regime as a whole, it is necessary to compare how the models represent the large-scale dynamics.

\begin{figure*}
\centering 
\subfloat[Results for the Ben~1 simulations]{
\label{fig:sub-first_wind_zonal_mean}
\includegraphics[width=1.0\linewidth]{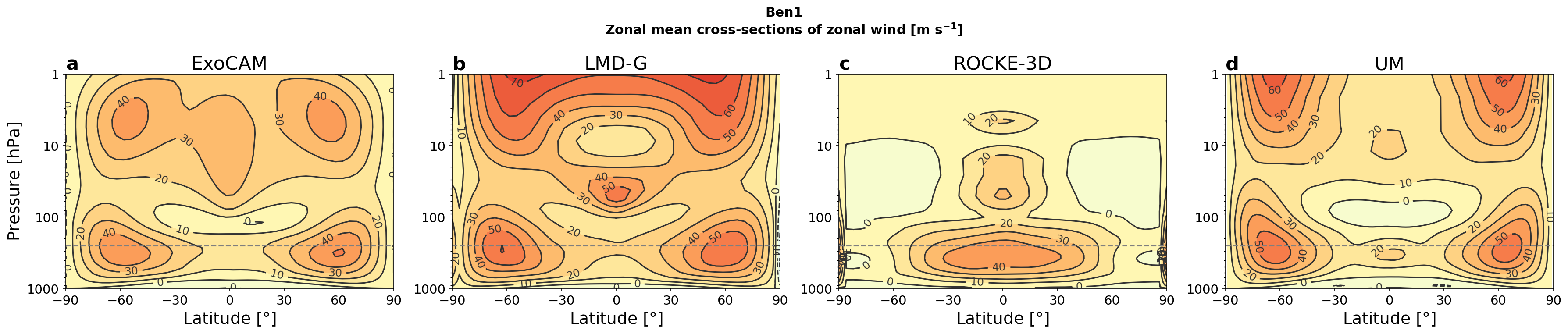}}
\\
\subfloat[Results for the Ben~2 simulations]{
\label{fig:sub-second_wind_zonal_mean}
\includegraphics[width=1.0\linewidth]{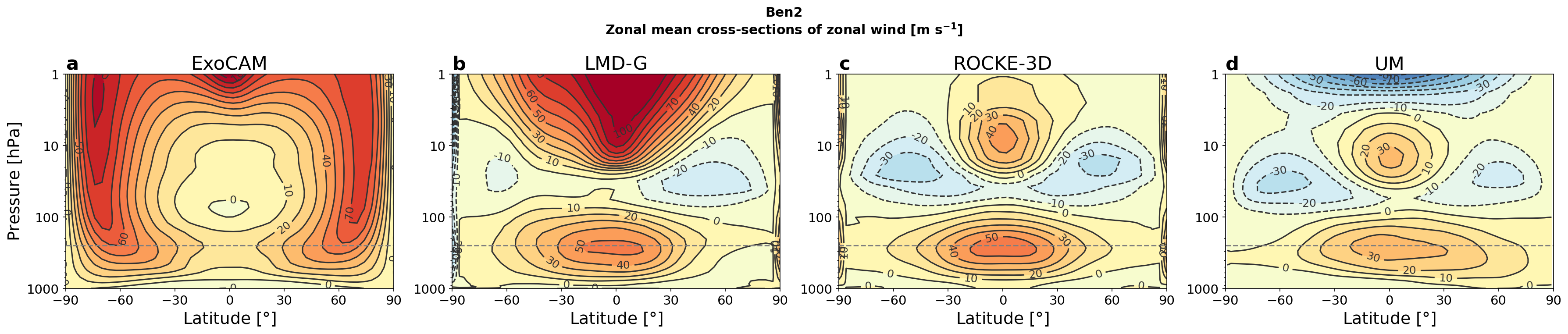}}
\caption{Vertical cross-sections of the zonal mean zonal wind (\si{\m\per\s}). The gray dashed horizontal line marks the \SI{250}{\hecto\pascal} level used in Fig.~\ref{fig:fig_substellar_Hdecomposition_Ben1} and Fig.~\ref{fig:fig_substellar_Hdecomposition_Ben2}. 
\label{fig:zm_u}}
\end{figure*}

\begin{figure*}
\centering 
\subfloat[Results for the Ben~1 simulations]{
\label{fig:sub-first_TLSF}
\includegraphics[width=1.0\linewidth]{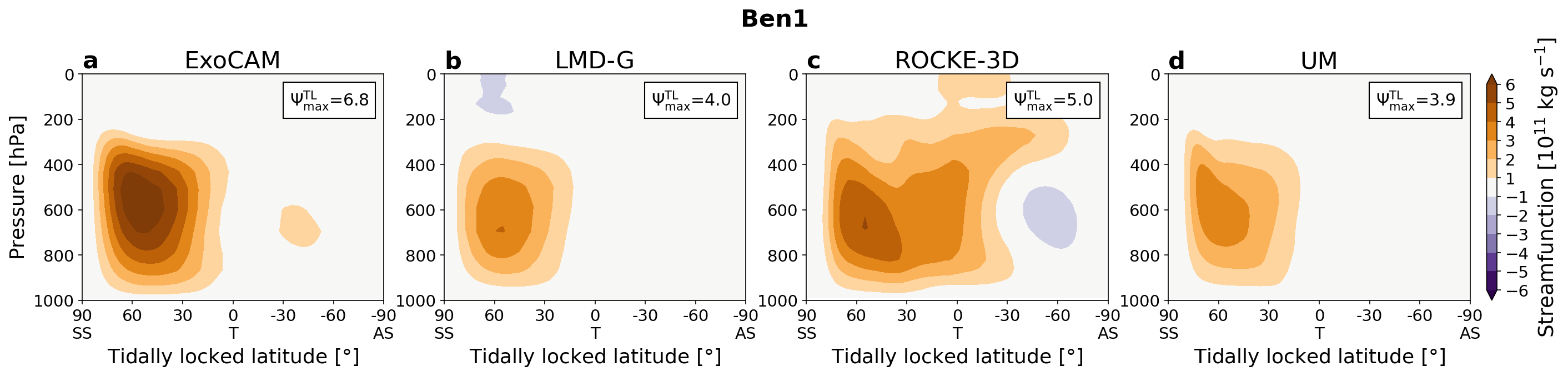}}
\\
\subfloat[Results for the Ben~2 simulations]{
\label{fig:sub-second_TLSF}
\includegraphics[width=1.0\linewidth]{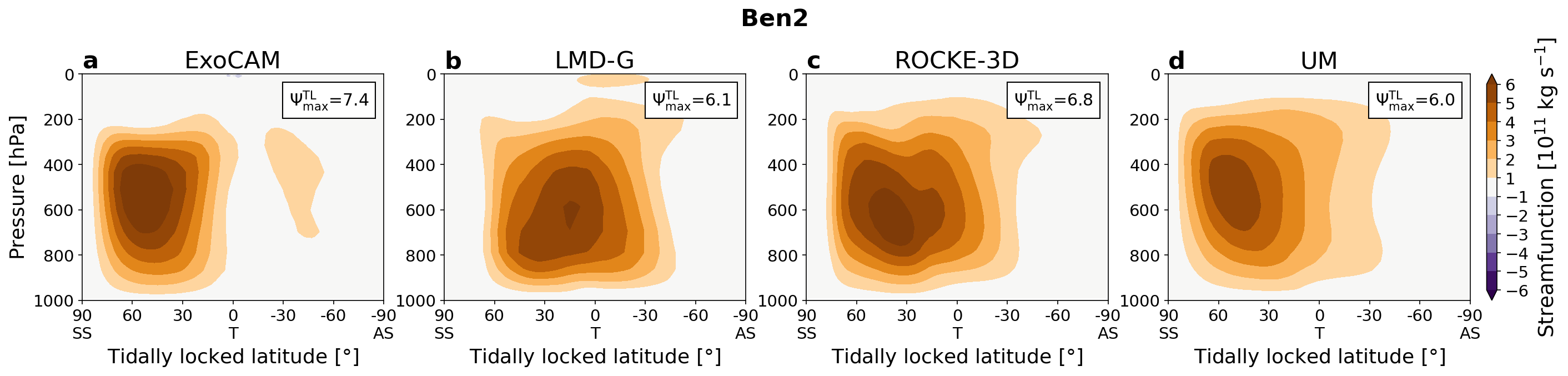}}
\caption{Overturning circulation shown by the tidally locked mass streamfunction $\Psi^{TL}$ (10$^{11}$~\si{\kg\per\s}). The mass flux is clockwise where $\Psi^{TL}$ is positive and anticlockwise where $\Psi^{TL}$ is negative. The maximum of $\Psi^{TL}$ in each GCM is shown in the upper left corner. Substellar point (SS) is at \ang{90} latitude, antistellar point (AS) is at \ang{-90} latitude. 
\label{fig:TLSF}}
\end{figure*}

Figure~\ref{fig:fig_tsurf} shows the surface temperature maps and the winds at the \SI{250}{\hecto\pascal} isobaric surface for the Ben~1 (panel a) and Ben~2 (panel b) cases.  First we can see that the inter-model spread in the global mean surface temperature is \SI{7}{\K} for the Ben~1 and \SI{6}{\K} for the Ben~2 simulations (see also Table~\ref{tab:glob_diag}). In both cases ExoCAM is on average the warmest and the UM the coldest, similar to what is found for the Hab~1 case \citep{Sergeev21_THAI}. For the Ben~1 case the day-night thermal contrast is the largest for ROCKE-3D, followed by the UM, LMD-G and ExoCAM. In the case of ROCKE-3D two wind gyres form in the western hemisphere at high latitudes, in the coldest regions (see Fig.~\ref{fig:fig_tsurf}b, bottom left plot). Those gyres are a manifestation of a Rhines rotator dynamical regime, characteristic for slow-rotating tidally locked exoplanet simulations \citep[e.g.][]{carone2015connecting,Noda17_circulation,Haqq_Misra2018,hammond_2020,WangYang2021} ; this is a regime where a strong upper atmosphere super-rotation is combined with a strong upwelling beneath the substellar point. 
The other three GCM simulations produce a fast-rotator circulation regime: without the gyres, but with two extratropical zonal jets. This is best illustrated by the lower part of Fig.~\ref{fig:zm_u} which shows the vertical cross-sections of the zonal mean zonal wind.
Similar to the findings of \citet{Sergeev2020} for TRAPPIST-1e, the thermal contrast between the day side and the night side is smaller in this regime than in the slow-rotator regime. This is not only important for TRAPPIST-1e but generally for any tidally-locked rocky exoplanet with short enough orbital periods,  i.e. short enough for the Rossby deformation radius to be comparable with the planet’s radius \citep{carone2015connecting,carone2016connecting,Haqq_Misra2018,Penn2018,Pierrehumbert2019,Zhang2020review}.

In the Ben~2 case, the circulation regime for two of the models (LMD-G and the UM) changes to the Rhines rotator (Fig.~\ref{fig:fig_tsurf}b). The day-night contrast stays the largest in ROCKE-3D's simulation, followed then by that in ExoCAM.
Despite the circulation regime change, LMD-G and the UM still have a lower day-night temperature contrast than that in ExoCAM.
Note, however, that the magnitude of this contrast is overall smaller than in the Ben~1 case due to overall more efficient heat redistribution (see Fig.~\ref{fig:TLSF}).

\begin{figure*}
\centering 
\includegraphics[width=1.0\linewidth]{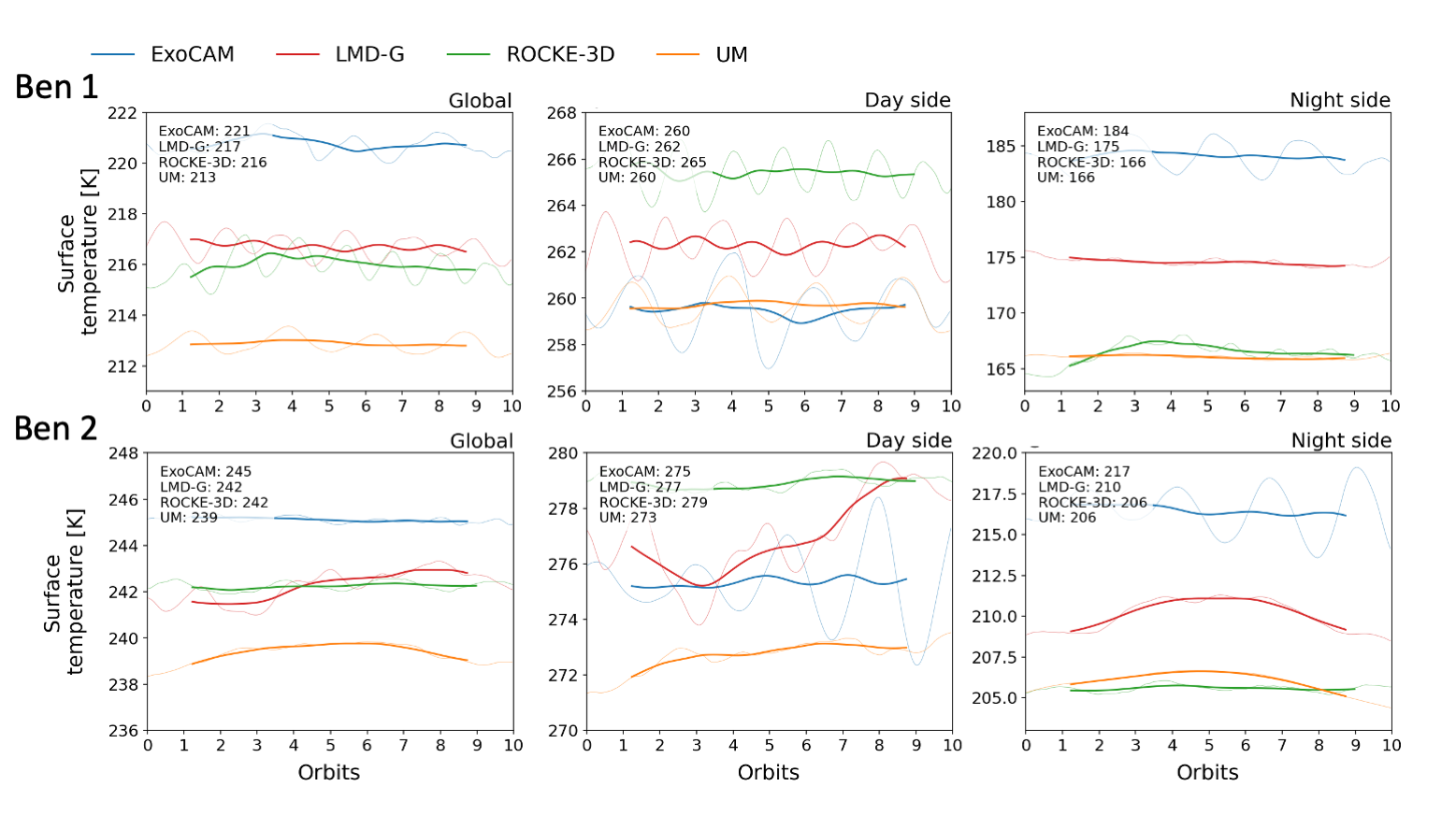}
\caption{Temporal variability of the surface temperatures (K) for the Ben~1 (top row) and Ben~2 (bottom row) simulations. Each column shows global mean, day-side mean and night-side mean. Thin lines show the raw data, thick lines show a 2.5-orbit rolling mean.
\label{fig:variability}}
\end{figure*}

Ben~1 and Ben~2 simulations both exhibit small temporal variability, as illustrated for surface temperatures in Fig.~\ref{fig:variability}, due most likely to the manifestation of weather patterns. It has indeed already been shown that on synchronous planets (with a fixed solar forcing), a temporal variability is nevertheless expected \citep{Joshi1997,Merlis2010,Ding2021}. The four GCMs show a similar small level of global mean surface temperature high-frequency variability ($\sim$~1~K) for Ben~1 and Ben~2 simulations. However, on a period of 10 consecutive orbits, the frequency of temperature oscillations varies from one model to another. The ExoCAM simulations show much greater night-side (Fig.~\ref{fig:variability}, right) and day-side (Fig.~\ref{fig:variability}, middle) variability than the other GCMs, but these opposite-phase oscillations offset each other on the global mean (Fig.~\ref{fig:variability}, left). This behavior, which is also pronounced in the ExoCAM Hab~1 and Hab~2 simulations (Part 2, \citealt{Sergeev21_THAI}), results from periodic weather systems traveling around the planet. Unfortunately, the design of the Ben~1 and Ben~2 cases of the THAI protocol \citep{Fauchez2020THAI} -- which limited the output files to 10 orbits --  do not allow to probe the low-frequency temporal variability, but there are several indications that it exists, as suggested by rolling averages in Fig.~\ref{fig:variability}. This is also indicated by the presence of north-south asymmetries in the different mean values of the models (see e.g., Fig.~\ref{fig:fig_tsurf} and Fig.~\ref{fig:zm_u}), somewhat similar to the results of \citet{Noda17_circulation}, Section 4.3. There do not appear to be clear correlations between the different model variabilities and the land properties adopted in the models (especially the number and depth of layers ; see Table~\ref{tab:models}, rightmost column) over the time range explored from the THAI simulations. We leave the question of why weather systems seem to be more pronounced on ExoCAM than other GCMs for future studies. We also recall that the level of temporal variability of the different models is far below the detectability threshold of JWST and the astronomical observatories of the coming decades (Part 3, \citealt{Fauchez21_THAI}).

The fact that the models reside in different rotational regimes is likely because the radius and rotation period of TRAPPIST-1e places it just at the edge between fast and Rhines (intermediate) rotation regimes \citep{carone2015connecting,carone2016connecting,Haqq_Misra2018}, as already identified in \citet{Sergeev2020} and \citet{Carone2018} specifically for TRAPPIST-1e. As a matter of fact, \citet{Sergeev2020} showed that just a change in the parameterization of convection can drive a  GCM into one or the other dynamical regime. Unlike the simulations in \citet{Edson11}, initial conditions are unlikely to play a role here as all simulations start from isothermal conditions of 300~K and null winds. We also identified in preliminary THAI simulations (of the UM model) that subtle changes (here, of correlated-k tables) are able to settle the GCM to a different rotation regime.
We also note that a similar behaviour appears in the aquaplanet THAI simulations, with ROCKE-3D being an outlier and settling in a fast-rotator regime \citep{Sergeev21_THAI}. 
Changes in surface friction can also drive GCMs into one or the other climate regime \citep{carone2016connecting}, but we caution that the surface frictions used in \citet{carone2016connecting} may be unrealistically high thus potentially overestimating the importance of this effect \citep{Wicker:2022}. More generally, this shows that the tipping point is highly non-linear and that small parameter changes can lead to a different climate regime.

Similar regime changes are seen between N$_2$ and CO$_2$-rich atmospheres in the cases including moisture \citep{Sergeev21_THAI}, and it is apparent that the simulations are sensitive to several parameters for this regime. Model parameterizations affect the eddy generation in our GCMs in different ways, which may then tip the balance one way or another. Further work has been performed to explore the potential bistability of this climate state (Sergeev et al., submitted).

\begin{figure*}
\centering 
\includegraphics[width=0.9\linewidth]{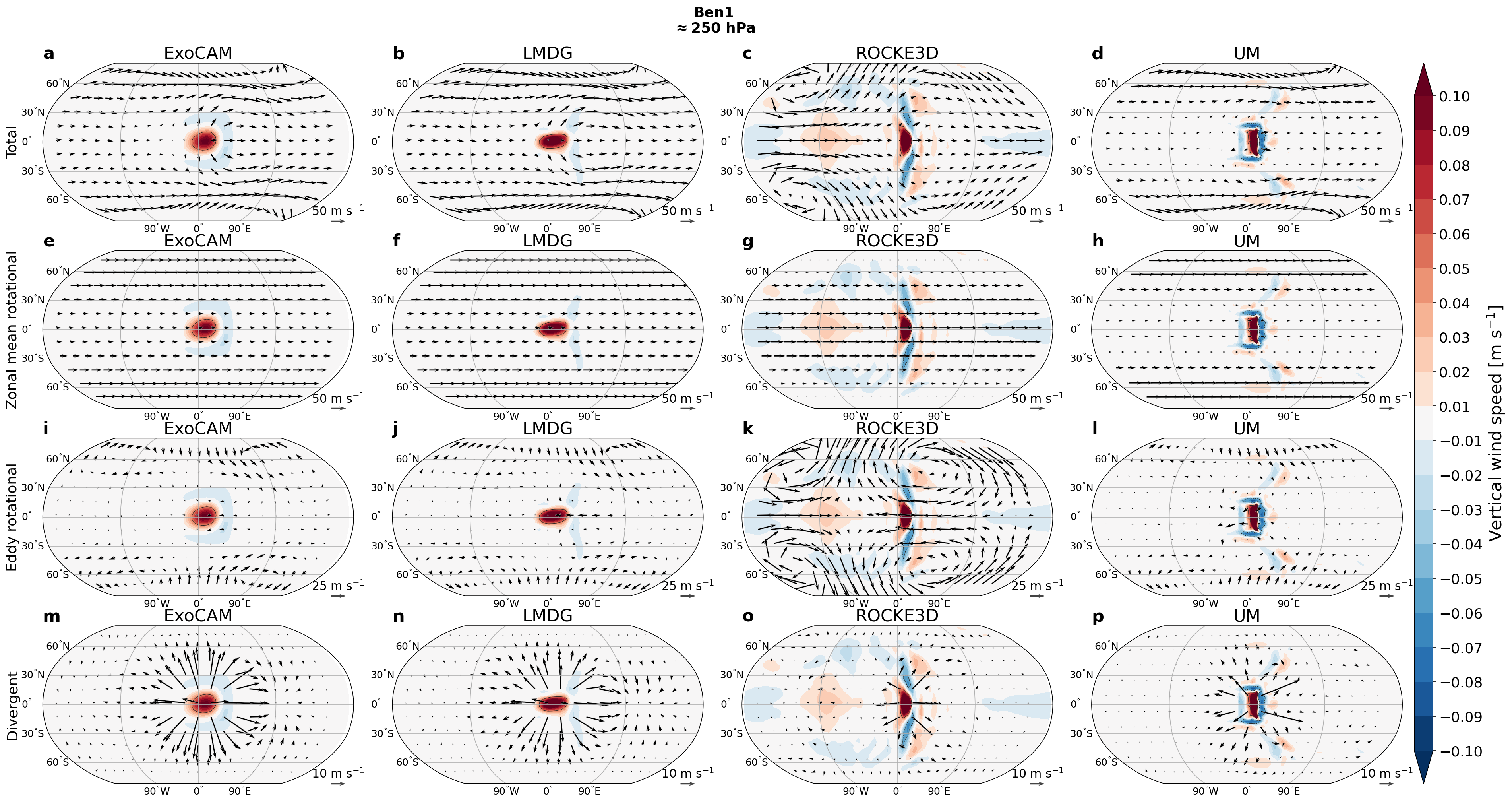}
\caption{Helmholtz decomposition of the horizontal wind at \SI{250}{\hecto\pascal} in the Ben~1 cases (quivers): (a--d) total wind, (e--h) zonal mean rotational component, (i--l) eddy rotational component, (m--p) divergent component. Note the different scaling of the eddy rotational and divergent components. Also shown is the upward wind velocity (shading, \si{\m\per\s}) with the \SI{0.05}{\m\per\s} highlighted by a black contour. Note that for ExoCAM, only the pressure velocity ($\omega$, \si{\pascal\per\s}) is available in the output, so the vertical velocity ($w$, \si{\m\per\s}) is approximated as $w=-\omega/\rho g$, where $\rho$ is air density and $g$ is the acceleration due to gravity.
\label{fig:fig_substellar_Hdecomposition_Ben1}}
\end{figure*}

\begin{figure*}
\centering 
\includegraphics[width=0.9\linewidth]{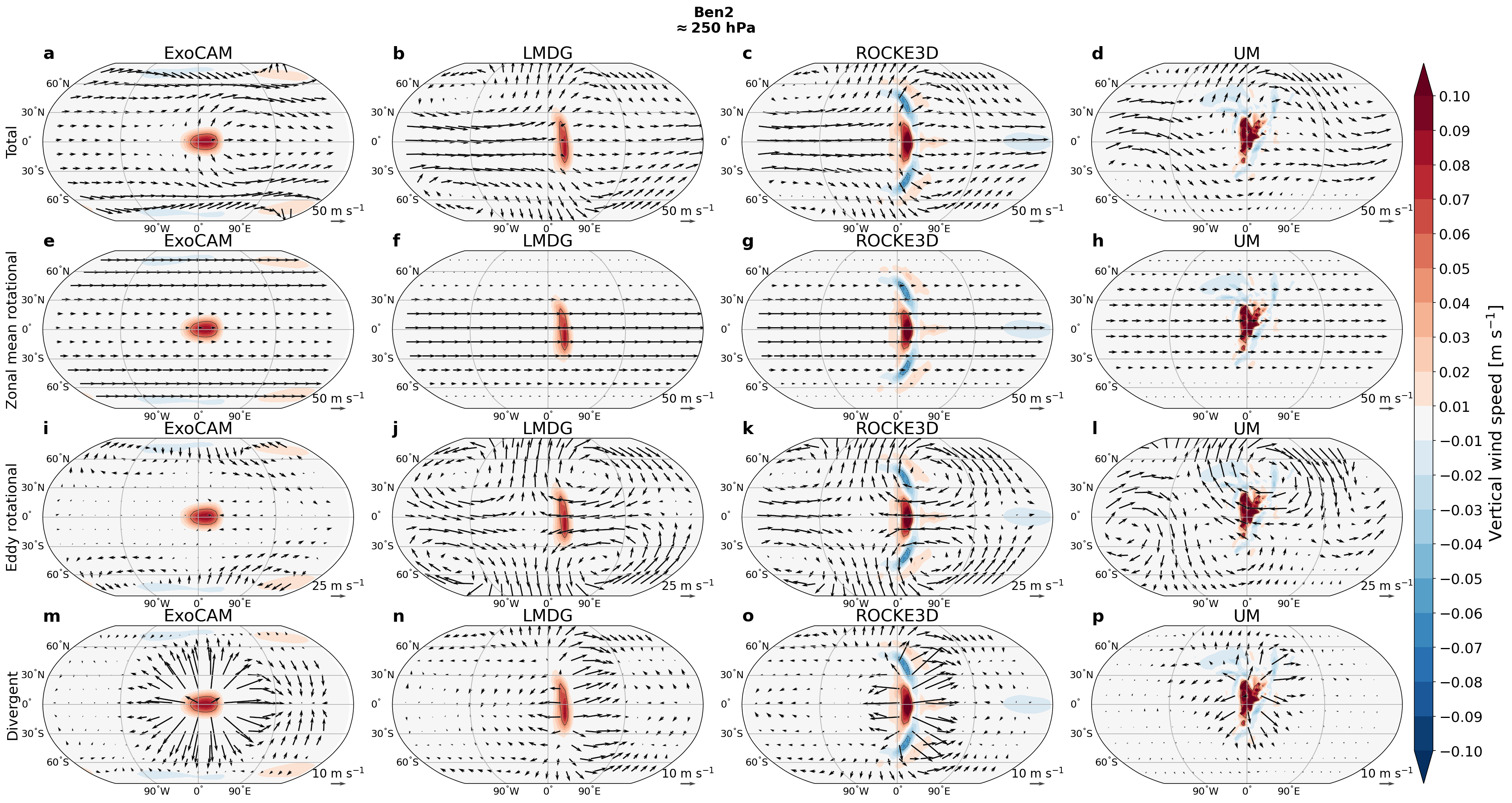}
\caption{Same as Fig.~\ref{fig:fig_substellar_Hdecomposition_Ben2} for the Ben~2 simulations.
\label{fig:fig_substellar_Hdecomposition_Ben2}}
\end{figure*}

To get a more detailed view of these dynamical regimes, we used the Helmholtz decomposition methodology of \citet{Hammond21}. The total wind flow at 250~hPa is decomposed into its zonal mean rotational, eddy rotational and divergent flows, as can be seen in Figs.~\ref{fig:fig_substellar_Hdecomposition_Ben1} and \ref{fig:fig_substellar_Hdecomposition_Ben2} for the Ben~1 and Ben~2 simulations, respectively. The zonal means of the rotational wind clearly show the prograde jets induced by the fast rotation of TRAPPIST-1e (Fig.~\ref{fig:fig_substellar_Hdecomposition_Ben1}e--h for the Ben~1 case and Fig.~\ref{fig:fig_substellar_Hdecomposition_Ben2}e--h for the Ben~2 case). 
The eddy component (as the deviation from the zonal mean) reveals planetary-scale stationary waves straddling the substellar longitude (Fig.~\ref{fig:fig_substellar_Hdecomposition_Ben1}i--l for Ben~1 and Fig.~\ref{fig:fig_substellar_Hdecomposition_Ben2}i--l for Ben~2).
For the Ben~1 simulations, only ROCKE-3D displays stationary gyres at mid-latitudes. 
As ExoCAM, LMD-G and the UM do not show those gyres, this is likely a sign of a faster atmospheric super-rotation placing them in the fast rotator regime. For the Ben~2 simulations, LMD-G, ROCKE-3D and the UM display stationary gyres at mid-latitudes. Only ExoCAM produces the fast-rotating regime, with no mid-latitudes gyres.
The asymmetry visible in the UM simulation of the Ben~2 case is likely an artifact of a relatively short simulation length.

These differences in circulation regime are also visible in the tidally locked mass streamfunction (Fig.~\ref{fig:TLSF}), which describes the overturning circulation as a function of the substellar latitude (between \ang{90} at the substellar and \ang{-90} at the anti-substellar points). 
In all eight simulations (Ben~1 and 2, and for the four GCMs), the mass streamfunction $\Psi^{TL}$ shows a single circulation cell between the day side and the night side (with air parcels rising at the substellar point and subsiding on the night side). There is some variability in the strength of the circulation across models, but the circulation is consistently stronger in the Ben~2 than in the Ben~1 simulations. This is likely due to the fact that the intensity of convection is stronger at the substellar point in Ben~2 (due mostly to the fact that longwave CO$_2$ absorption by the dayside atmosphere leads to surface warming that further increases convection), which is further strengthened by the strong radiative nightside cooling (due to stronger longwave emission by CO$_2$). This reflects directly on Fig~\ref{fig:fig_OLR}, which shows that the OLR is weaker on the dayside and stronger on the nightside in the Ben2 simulations than in the Ben1 simulations.
We observe that the circulation cells extend to higher substellar latitudes for simulations trapped in the Rhines rotation regime (ROCKE-3D for Ben~1 ; LMD-G, ROCKE-3D and the UM for Ben~2) than those trapped in the fast rotation regime, as previously identified in \cite{Haqq_Misra2018}.

\medskip

Last but not least, we identify that the main source of differences between the models appears in the upper part of the atmosphere. In Figs.~\ref{fig:fig_substellar_Ben1} and \ref{fig:fig_substellar_Ben2}, temperature differences between models peak in the upper layers of the atmosphere reaching about \SIrange{10}{20}{\K} in both the Ben~1 and Ben~2 simulations. This is also where variability intrinsic to the models (visible in shade in Figs.~\ref{fig:fig_substellar_Ben1} and Fig.~\ref{fig:fig_substellar_Ben2}) is the largest.

In Fig.~\ref{fig:zm_u}, wind distribution is also very different from one model to another above the troposphere. Simulations in the fast-rotator regime (all but ROCKE-3D in the Ben~1 case; ExoCAM in the Ben~2 case) all exhibit two extratropical high-altitude (near \SIrange{1}{10}{\hecto\pascal}) prograde jets. In ExoCAM's Ben~2 simulation, these jets extend to the troposphere and merge with the two aforementioned tropospheric jets. Simulations in the Rhines rotator regime (ROCKE-3D in the Ben~1 case; all but ExoCAM in the Ben~2 case) all exhibit an equatorial (near \SIrange{1}{10}{\hecto\pascal}) prograde jet and two high-latitude mid-altitude (near \SIrange{10}{100}{\hecto\pascal}) retrograde jets.

While we do not show wind patterns above \SI{1}{\hecto\pascal} in the atmosphere here, because some of the models (the UM, specifically) have a model top that stops just above this pressure, we do observe strong discrepancies near \SI{1}{\hecto\pascal} and above. These discrepancies can originate from differences in radiative forcings (see e.g. Fig.~\ref{fig:fig_substellar_Ben1}C and \ref{fig:fig_substellar_Ben2}C), differences in the location and spacing of vertical levels (this choice was left free in the THAI protocol), but most likely from differences in numerical damping parameterization (e.g. by the presence and formulation of a sponge layer). As detailed in the Methods section, each model uses its own recipe to simulate wind braking (as well as dissipation to smaller scales) with varying degrees of amplitude and levels of sophistication. For example, winds above \SI{1}{\hecto\pascal} are close to zero in ROCKE-3D simulations (for both Ben~1 and Ben~2 simulations), which comes from the presence of enhanced Rayleigh damping in the topmost layers of the model; while LMD-G exhibits strong, super-rotating winds above \SI{1}{\hecto\pascal} (for both Ben~1 and Ben~2 simulations), which is made possible by the absence of a sponge layer. Wind braking has indeed already been shown to have a significant impact on the stratospheric circulation \citep{Carone2018}. Additional sensitivity studies performed with ROCKE-3D using moderate Rayleigh damping (the default ROCKE-3D sponge layer configuration) re-establishes a high-altitude stratospheric super-rotating jet. These sensitivity studies also reveal that the impact of the upper atmosphere damping parameterization on the climate state of the lower atmosphere is very weak (e.g., mean surface temperature only differs by 0.1K).

Although the direct impact on the lower atmosphere and the surface of the presence (or absence) of a super-rotating jet appears to be limited, this could have an indirect impact by affecting the photochemistry and the distribution of aerosols in the stratosphere, which could also impact the observations of the planet \citep{Chen2019,Boutle2020}. There is also a causality in the opposite direction: the radiative heating due to different gases drives the atmospheric circulation in the upper atmosphere (and may be one of the cause of inter-model differences). Given the proximity of these dynamical features to the top of our model domains, and the additional processes which become important at low pressures, we reserve a full analysis of this element to a future work, where the additional model treatments can be included and maximum altitude raised.

\section{Conclusions} \label{sec:conclusions}

In this manuscript, we reported the results of the first part of the THAI project, which compares simulation results of four state-of-the-art 3D Global Climate Models (ExoCAM, LMD-Generic, ROCKE-3D, the Unified Model) for the exoplanet TRAPPIST-1e. This first part consisted of comparing and analysing the results of the so-called `Ben~1' and `Ben~2' cases, which assume that the planet has a dry surface and atmosphere, with N$_2$-dominated and CO$_2$-dominated, respectively. 

These simulations were designed as benchmarks to test how radiative processes, small-scale dynamics (dry turbulence and convection) and large-scale dynamics in 3D GCMs impact the climate of TRAPPIST-1e, whether the GCMs agree with each other and what implications might the inter-model differences have for observable features of this planet.

To first order, the four models give results in fairly good agreement. The inter-model spread in the global mean surface temperature amounts to \SI{7}{\K} for the N$_2$-dominated atmospheric composition and \SI{6}{\K} for the CO$_2$-dominated one. The radiative fluxes (both stellar and thermal) are also remarkably similar (variations of radiative heating rates between the four models of always less than \SI{5}{\percent}), from the surface (\SI{\approx 1000}{\hecto\pascal}) up to atmospheric pressures \SI{\approx 5}{\hecto\pascal}.

Moderate differences between the models appear in the atmospheric circulation pattern (winds) and the (stratospheric) thermal structure. These differences arise between the models from (1) large scale dynamics because TRAPPIST-1e lies at the tipping point between two different circulation regimes (fast and Rhines rotators) in which the models can be alternatively trapped ; and (2) parameterizations used in the upper atmosphere such as numerical damping (e.g., the presence and formulation of a sponge layer). 

Transit spectra (computed in THAI part~3 ; see \citealt{Fauchez21_THAI}) are shown to be sensitive to atmospheric pressures down to \num{e0} and \SI{e-2}{\pascal} for the Ben~1 \& Ben~2 cases, respectively, in the core of the 4.3~$\mu$m CO$_2$ band \citep{Fauchez21_THAI}, and considering the typical spectral resolution of the instruments (R~$\sim$~10$^2$-10$^3$) onboard JWST. At the typical spectral resolution of ground-based spectrographs (R~$\sim$~10$^5$), transit spectra are found to be sensitive to atmospheric pressures down to \num{e-3} and \SI{e-7}{\pascal} for the Ben~1 \& Ben~2 cases, respectively, in the core of CO$_2$ lines (of the 4.3~$\mu$m CO$_2$ band). Efforts should thus be made to improve the description of the processes affecting the thermal structures of planets above \SI{1}{\hecto\pascal} or so in GCMs, because temperature-induced variations of the stratospheric scale height can significantly alter the amplitude of the strongest molecular absorption bands \citep{Fauchez21_THAI}).

Overall, the main objective of these dry, benchmark simulations has been achieved. The simulations demonstrate that the main differences between the  models presented for the aquaplanet simulations \citep[described in Part II of the THAI trilogy, see][]{Sergeev21_THAI} come mostly from water vapor and clouds (although differences due to convection and RT are amplified when water vapor and clouds are present). Similarly, most of the differences in the simulated transit spectra of TRAPPIST-1e \citep[described in Part III of the THAI trilogy, see][]{Fauchez21_THAI} are dominated by the variations in the 3D distribution of water vapor and clouds.

\acknowledgments
This project has received funding from the European Union's Horizon 2020 research and innovation program under the Marie Sklodowska-Curie Grant Agreement No. 832738/ESCAPE.
M.T. thanks the Gruber Foundation for its generous support to this research. M.T. acknowledges support from the PORTAL BRAIN-be 2.0 BELSPO project. M.T. thanks Ray Pierrehumbert and Jeremy Leconte for useful discussions on this work. M.T. thanks Shuang Wang for helping to track a mistake in the original version of the manuscript. M.T. was granted access to the High-Performance Computing (HPC) resources of Centre Informatique National de l'Enseignement Sup\'erieur (CINES) under the allocations \textnumero~A0020101167 and A0040110391 made by Grand \'Equipement National de Calcul Intensif (GENCI).
This work has been carried out within the framework of the National Centre of Competence in Research PlanetS supported by the Swiss National Science Foundation.
M.T. acknowledges the financial support of the SNSF.\\
M.T. and F.F. thank the LMD Generic Global Climate team for the teamwork development and improvement of the model.
T.F. and M.J.W. acknowledge support from the GSFC Sellers Exoplanet Environments Collaboration (SEEC), which is funded in part by the NASA Planetary Science Divisions Internal Scientist Funding Model.
M.J.W was supported by NASA's Nexus for Exoplanet System Science (NExSS). Resources supporting the ROCKE-3D simulations were provided by the NASA High-End Computing (HEC) Program through the NASA Center for Climate Simulation (NCCS) at Goddard Space Flight Center.
D.E.S., I.A.B., F.H.L, J.M. and N.J.M. acknowledge use of the Monsoon system, a collaborative facility supplied under the Joint Weather and Climate Research Programme, a strategic partnership between the Met Office and the Natural Environment Research Council.
We acknowledge support of the Met Office Academic Partnership secondment program.
This work was partly supported by a Science and Technology Facilities Council Consolidated Grant (ST/R000395/1), UKRI Future Leaders Fellowship (MR/T040866/1), and the Leverhulme Trust (RPG-2020-82).
J.H.M. acknowledges funding from the NASA Habitable Worlds program under award 80NSSC20K0230. \\
The authors thank the two anonymous reviewers whose comments helped to improve the quality and clarity of this manuscript.\\
The THAI GCM intercomparison team is grateful to the Anong's THAI Cuisine restaurant in Laramie for hosting its first meeting on November 15, 2017.\\
Numerical experiments performed for this study required the use of supercomputers, which are energy intensive facilities and thus have non-negligible greenhouse gas emissions associated with them.
We estimate that the final versions of Ben~1 $\&$ Ben~2 experiments resulted in \SI{\approx 840}{\kg}CO$_2$e, including \SI{\approx 750}{\kg} from ExoCAM runs, \SI{\approx 20}{\kg} from LMD-G, \SI{\approx 70}{\kg} from ROCKE-3D and \SI{\approx 3}{\kg} from the UM.

%

\software{
Scripts to process and visualize THAI data are available on GitHub: \url{https://github.com/projectcuisines/thai_trilogy_code} and are dependent on the following Python libraries: \texttt{aeolus} \citep{aeolus}, \texttt{iris} \citep{iris}, \texttt{jupyter} \citep{jupyter}, \texttt{matplotlib} \citep{Hunter2007}, \texttt{numpy} \citep{numpy}, \texttt{windspharm} \citep{windspharm}, \texttt{xarray} \citep{xarray}.
ExoCAM \citep{Wolf2015} is available on Github: https://github.com/storyofthewolf/ExoCAM.  The radiative transfer component of ExoCAM is available on Github: https://github.com/storyofthewolf/ExoRT. The National Center for Atmospheric Research Community Earth System Model version 1.2.1, a prerequisite for using ExoCAM, is available publicly here:  https://www.cesm.ucar.edu/models/cesm1.2/tags/.  The Met Office Unified Model is available for use under licence; see \url{http://www.metoffice.gov.uk/research/modelling-systems/unified-model} . ROCKE-3D is public domain software and available for download for free from \url{https://simplex.giss.nasa.gov/gcm/ROCKE-3D/}. Annual tutorials for new users take place annually, whose recordings are freely available on line at \url{https://www.youtube.com/user/NASAGISStv/playlists?view=50&sort=dd&shelf_id=15}. LMD-G is available upon request from Martin Turbet (martin.turbet@lmd.ipsl.fr) and François Forget (francois.forget@lmd.ipsl.fr).}



\appendix

\section{Data accessibility}
All our GCM THAI data are permanently available for download here: \url{https://ckan.emac.gsfc.nasa.gov/organization/thai}, with variables described for each dataset. If you use these data please cite the current paper and add the following statement: "THAI data have been obtained from \url{https://ckan.emac.gsfc.nasa.gov/organization/thai}, a data repository of the Sellers Exoplanet Environments Collaboration (SEEC), which is funded in part by the NASA Planetary Science Divisions Internal Scientist Funding Model."\\
Scripts to process the THAI data are available on GitHub: \url{https://github.com/projectcuisines}\\

\bibliography{biblio}{}
\bibliographystyle{aasjournal}



\end{document}